\documentclass[12pt]{article}
\input psfig.sty
\input epsfig.sty

\begin{document} \begin{titlepage}
\rightline{\vbox{\halign{&#\hfil\cr
&SLAC-PUB-8831\cr
&May 2001\cr}}}

\begin{center} 

\bigskip\bigskip\bigskip
{\Large\bf Indirect Signatures of $CP$ Violation in the Processes $\gamma \gamma \rightarrow \gamma \gamma$, $\gamma Z$, and $ZZ$}
\footnote{Work supported by the Department of 
Energy, Contract DE-AC03-76SF00515}

\medskip

\normalsize 
{\bf  JoAnne L. Hewett and Frank J. Petriello}
\vskip .3cm
Stanford Linear Accelerator Center \\
Stanford University \\
Stanford CA 94309, USA\\
\vskip .3cm
\medskip

\end{center} 

\begin{abstract}

This paper studies the utility of the processes $\gamma \gamma \rightarrow \gamma \gamma$, $\gamma Z$, and $ZZ$ in searching for sources of $CP$ violation arising from energy scales beyond the production thresholds of planned future colliders.  In the context of an effective Lagrangian approach we consider the most general set of $CP$ odd SU(2) $\times$ U(1) operators that give rise to genuinely quartic gauge boson couplings which can be probed in $2 \rightarrow 2$ scattering processes at a $\gamma \gamma $ collider.  We study each process in detail, emphasizing the complementary information that is obtained by varying the initial beam polarizations.  Finally, we compare our results to other constraints in the literature on $CP$ odd gauge boson interactions and quartic gauge boson couplings; the search reaches obtained here are typically stronger and nicely complement previous studies which have focused primarily on $W$ boson, top quark, or Higgs production.

\end{abstract}
\end{titlepage}

\newpage

\section{Introduction}

Future $e^+ e^- $ colliders will likely have the option of operating in $ \gamma \gamma $ or $ e \gamma $ collision modes~\cite{Russians}.  These modes are reached by Compton scattering laser light off one or more of the incoming fermion beams, and then colliding the resulting high energy photons with the remaining fermion beam or with each other.  There is a large potential for $ e \gamma$ and $\gamma \gamma$ collisions to elucidate possible physics beyond the Standard Model; previous investigations have focused on anomalous couplings~\cite{Anom}, searches for extra dimensions~\cite{ExtraD}, properties of supersymmetry~\cite{MSSM}, and a broad host of other topics.  

One subject that deserves further study is $CP$-violating gauge boson self-couplings.  $CP$ violation is one of the most poorly understood aspects of the Standard Model (SM).  Present data merely fixes the value of the $CP$ violating phase in the CKM matrix, and cannot test if this phase constitutes the only source for $CP$ violation.  In fact, studies of baryogenesis within the SM suggest that additional $CP$ violating terms are required in order to generate the observed baryon asymmetry in the universe~\cite{Baryo}.  Most discussions of $CP$ violation at photon colliders work within the context of models such as supersymmetry, and focus upon either Higgs or top quark production~\cite{MSSMCP}.  However, since almost every extension of the SM contains new $CP$ violating phases, it is desirable to eliminate any possible theoretical bias by studying $CP$ violation within the generic context of an effective Lagrangian approach, without assuming any underlying mechanisms.  A handful of such works have been performed in the past~\cite{CP,Choi}, concentrating on $ \gamma \gamma \rightarrow H$, $\gamma \gamma \rightarrow W^+ W^- $, or top quark production.  Here, we extend these studies by examining possible $CP$ violating quartic gauge boson couplings.  $\gamma \gamma$ colliders are particularly suited to studying such couplings; they can be probed in $2 \rightarrow 2$ scattering processes, unlike at $e^+ e^- $ colliders.

The SM contributions to the processes  $\gamma \gamma \rightarrow \gamma \gamma$, $\gamma \gamma \rightarrow \gamma Z$, and $\gamma \gamma \rightarrow ZZ$, including electroweak contributions, were first computed in~\cite{Jikia}.  Recently, they have been reexamined~\cite{Goun_gg1,Goun_gg2,Goun_gg3,Goun_gg4}, and shown to exhibit several interesting features that motivate our analysis.  The SM amplitudes vanish at tree level, and may therefore be quite sensitive to the effects of new physics.  At one loop they acquire large imaginary parts, completely dominated at high energies by the helicity amplitudes ${\cal M}_{\pm \pm \pm \pm}$, ${\cal M}_{\pm \mp \pm \mp}$, and ${\cal M}_{\pm \mp \mp \pm}$, where $ \pm$ denotes the direction of transverse polarization relative to the beam direction.  Such amplitudes can interfere strongly with $CP$ violating contributions from new sources, yielding greatly enhanced sensitivity to these new effects.  In addition, the ability to polarize both the initial laser light and fermion beams allows the construction of observables which are sensitive only to these interference effects.  These properties, together with the experimental cleanliness of $\gamma \gamma \rightarrow \gamma \gamma$, $\gamma Z$, and $ZZ$ scattering, suggest that these processes might provide powerful tools in searches for new physics.

In this paper we will show that $\gamma \gamma \rightarrow \gamma \gamma$, $\gamma Z$, and $ZZ$ at a photon collider can provide sensitive tests of $CP$ violation in the gauge boson sector.  We limit our study to genuinely quartic gauge boson operators, as contributions to the two point functions are strongly excluded by current data and the three point functions are likely better tested elsewhere.  We consider manifestly SU(2) $\times$ U(1) invariant operators constructed from the appropriate field strength tensors, making no assumptions as to the origin of electroweak symmetry breaking.  As a result, our operators are of dimension eight.  Throughout the paper we will identify features of our analysis that might be useful for other new physics searches using these processes.

The paper is organized as follows.  We first present the density matrix formalism for photon-photon collisions in some detail, both for completeness and to motivate the expression for the ensemble average of the scattering amplitude, which will be used throughout our analysis.  We then briefly discuss the construction of the anomalous $CP$ violating operators we have studied.  Our results for the three processes considered are presented next, including discussions of the various initial laser and fermion beam polarizations.  Finally, we present our conclusions, including a comparison of our resulting sensitivity to these effects with others in the literature.

\section{$ \gamma \gamma$ Collisions}

In this section we present the density matrix formalism for photon-photon collisions.  Our discussion is similar to that found in the literature (see, for example,~\cite{Choi}), but is included here to provide motivation for eq. (9), which is used throughout our analysis.

Consider a photon moving along the $z$ axis; its polarization vectors for positive and negative helicities are given by
\begin{equation}
\epsilon^{\pm} = \frac{1}{\sqrt{2}} \left(0,\mp 1,-i, 0 \right).
\end{equation}
We will denote the photons corresponding to states with these polarizations by $ | \pm \rangle$.  The most general photon state can be written in this basis as
\begin{equation}
| \alpha, \phi \rangle = -{\rm cos}(\alpha) e^{i \phi} | + \rangle \, + \, {\rm sin}(\alpha) e^{-i \phi} | - \rangle \,\,,
\end{equation}
where $ 0 \leq \alpha \leq \pi/2 $ and $ -\pi \leq \phi \leq \pi $.  Writing this state in 4-vector form demonstrates that the choices $ \alpha = 0$ and $ \alpha = \pi/2 $ lead to circularly polarized states, while $ \alpha = \pi/4$ leads to a linearly polarized state.  The angle $ \phi$ describes the direction of linear polarization in the $x-y$ plane.  

The experimental setup in $ \gamma \gamma $ colliders has been described in detail elsewhere~\cite{Russians}; briefly, one Compton scatters laser photons off fermion beams, and then collides the backscattered photons.  When observing the scattering process $ \gamma \gamma \rightarrow X$, the energies or helicities of the incoming photons are not known; they will be statistically distributed, with the specific distribution being determined by the details of the Compton scattering process.  When measuring the differential cross section, one no longer determines $  \mid \langle \phi_{final} | {\cal M} | \phi_{initial} \rangle \mid^2$, but rather the ensemble average of $ {\cal M}$:
\begin{equation}
\mid {\cal M} \mid^{2}_{ens} = \sum_{i} w^{i} \mid \langle \phi_{final} | {\cal M} | \phi_{initial}^i \rangle \mid^2 \,\, ,
\end{equation}
where the sum over $i$ sums over all possible initial states weighted by $ w^{i}$, the probability of their occurrence.  We can write this in the form
\begin{equation}
\mid {\cal M} \mid^{2}_{ens} = \sum_{j,k} \langle \phi_{final} | {\cal M} | \lambda_{j} \rangle  \langle \lambda_{k} | {\cal M} | \phi_{final} \rangle \, \rho_{jk} \,\, ,
\end{equation}
where we have introduced the density matrix for the incoming photons in the basis defined by $ \lambda_{j} $:
\begin{equation}
\rho_{jk} = \sum_{i} w^{i} \langle \lambda_{j} | \phi_{in}^i \rangle \langle \phi_{in}^i | \lambda_{k} \rangle \,\, .
\end{equation}
Let us compute the density matrix for the laser photon of eq. (2); introducing the notation $P_c = {\rm cos}(2 \alpha)$ and $ P_t = {\rm sin}(2 \alpha)$, we have
\begin{equation}
\rho = \frac{1}{2} \, \pmatrix{ 1 + P_c  & -P_t \, e^{2i \phi} \cr
-P_t \, e^{-2i \phi} & 1-P_c } \, .
\end{equation}
$P_c$ and $P_t$ measure the amounts of circular and linear polarization, respectively.  This is consistent with the polarizations for the values $\alpha = \pi/2$ and $ \alpha = \pi/4$ noted before.  After the Compton scattering process, the degrees of linear and circular polarization are described by distributions which are dependent upon the fraction of the fermion beam energy the laser photons acquire.  Denoting this fraction by $x$, and the helicity distribution functions by $\xi_{c,t} (x)$, the density matrix becomes
\begin{equation}
\rho = \frac{1}{2} \, \pmatrix{ 1+ \xi_c (x) & - \xi_t (x) e^{2i \phi} \cr
- \xi_t (x) e^{-2i \phi} & 1-\xi_c (x) } \, .
\end{equation}
The eigenvalues of this matrix are 
\begin{equation}
\mu_{ \pm} = 1 \pm \sqrt{ \xi_{t}^{2} + \xi_{c}^{2} } \, ;
\end{equation}
we no longer have a pure state unless $ \xi_{t}^{2} + \xi_{c}^{2} =1$.  

As the processes considered in this paper involve two Compton scattered photons, the complete density matrix will be the tensor product of the density matrix for each photon.  Also, since one of the initial laser photons will be moving along the $-z$ axis, we must take $ \phi \rightarrow - \phi$ in its density matrix.  Referring back to eqs. (4) and (7), we can write the ensemble average for the process $ \gamma \gamma \rightarrow X$ as
\begin{eqnarray}
\mid {\cal M} \mid^{2}_{ens} &=& \frac{1}{4} \,\, \bigg\{ \mid {\cal M}_{++} \mid^2 + \mid {\cal M}_{--} \mid^2 + \mid {\cal M}_{+-} \mid^2 + \mid {\cal M}_{-+} \mid^2 \nonumber \\ & & + \xi_c (x_1) \left[  \mid {\cal M}_{++} \mid^2 - \mid {\cal M}_{--} \mid^2 + \mid {\cal M}_{+-} \mid^2 - \mid {\cal M}_{-+} \mid^2 \right] \nonumber \\  & & + \xi_c (x_2) \left[  \mid {\cal M}_{++} \mid^2 - \mid {\cal M}_{--} \mid^2 - \mid {\cal M}_{+-} \mid^2 + \mid {\cal M}_{-+} \mid^2 \right] \nonumber \\  & & + \xi_c (x_1) \xi_c (x_2) \left[  \mid {\cal M}_{++} \mid^2 + \mid {\cal M}_{--} \mid^2 - \mid {\cal M}_{+-} \mid^2 - \mid {\cal M}_{-+} \mid^2 \right] \nonumber \\ & & + 2 \xi_{t} (x_1) \xi_{t} (x_2) \, {\rm Re}  \left[ {\cal M}_{+-} {\cal M}^{*}_{-+} e^{2i( \phi_1 + \phi_2)} + {\cal M}_{++} {\cal M}^{*}_{--}  e^{2i( \phi_1 - \phi_2)} \right] \nonumber \\ & & -2 \xi_{t} (x_1) \, {\rm Re} \left[ {\cal M}_{++} {\cal M}^{*}_{-+} e^{2i \phi_1} + {\cal M}_{+-} {\cal M}^{*}_{--} e^{2i \phi_1} \right] \nonumber \\ & &  -2 \xi_{t} (x_2) \, {\rm Re} \left[ {\cal M}_{+-} {\cal M}^{*}_{++} e^{2i \phi_2} + {\cal M}_{--} {\cal M}^{*}_{-+} e^{2i \phi_2} \right] \nonumber \\ & & + 2 \xi_{t} (x_1) \xi_c (x_2) \, {\rm Re} \left[ {\cal M}_{+-} {\cal M}^{*}_{--} e^{2i \phi_1} - {\cal M}_{++} {\cal M}^{*}_{-+} e^{2i \phi_1} \right] \nonumber \\ & &
+ 2 \xi_{c} (x_1) \xi_t (x_2) \, {\rm Re} \left[ {\cal M}_{--} {\cal M}^{*}_{-+} e^{2i \phi_2} - {\cal M}_{+-} {\cal M}^{*}_{++} e^{2i \phi_2} \right] \bigg\} \,\, ,
\end{eqnarray}
where, for example, ${\cal M}_{+-}$ denotes the helicity amplitude for incoming photons with helicities $+1$ and $-1$.  We have suppressed the final state $X$; depending upon the process considered the observable final state might require a sum over various helicities.  This expression will be used to construct observables for all the processes considered in this paper.

The physical cross section involves a convolution of this scattering amplitude with the photon number density function of each photon.  The explicit form of this function, as well as the forms for the linear and circular helicity distribution functions, can be found in the Appendix.  

\section{Constructing $CP$ Violating Operators}

Here we construct the most general set of operators that contribute to neutral gauge boson self-interactions, subject to the following constraints.  We consider only SU(2) $\times$ U(1) invariant operators, and further restrict these to $CP$-odd terms.  As the effects of $CP$-odd trilinear operators have been extensivley studied~\cite{Trilinear}, we consider only those terms that lead to quartic or higher gauge boson interactions.  We also make no assumption as to the mechanism of electroweak symmetry breaking.  These restrictions lead us to the following set of seven dimension eight operators constructed from the $W$ and $B$ field strength tensors:
\begin{eqnarray}
& & {\cal O}_{(BB)(BB)}=(B_{\mu \nu}B^{\mu \nu})(B^{\rho \sigma} \stackrel{\sim}{B}_{\rho \sigma} ), \,\,\,\, 
{\cal O}_{(WW)(WW)}=(W_{a \mu \nu}W_{a}^{\mu \nu})(W_{b}^{\rho \sigma} \stackrel{\sim}{W}_{b \rho \sigma} ),
\nonumber \\
& & {\cal O}_{(BB)(WW)}=(B_{\mu \nu}B^{\mu \nu})(W_{a}^{\rho \sigma} \stackrel{\sim}{W}_{a \rho \sigma} ), \,\,\,\,
{\cal O}_{(WW)(BB)}=(W_{a \mu \nu}W_{a}^{\mu \nu})(B^{\rho \sigma} \stackrel{\sim}{B}_{\rho \sigma} ), \nonumber \\
& & {\cal O}_{(BW)(BW)}=(B_{\mu \nu}W_{a}^{\mu \nu})(B^{\rho \sigma}\stackrel{\sim}{W}_{a \rho \sigma} ), \,\,\,\,
{\cal O}_{WBWB}=W_{a}^{\mu \nu}B_{\nu \rho}W_{a}^{\rho \sigma} \stackrel{\sim}{B}_{\sigma \mu} , \nonumber \\
& & {\cal O}_{BWBW}=B^{\mu \nu}W_{a \nu \rho}B^{\rho \sigma} \stackrel{\sim}{W}_{a \sigma \mu} \,\, .
\end{eqnarray}
Here $W_{a}^{\mu \nu}$ is the SU(2) field strength tensor and $B^{\mu \nu}$ the U(1) field strength tensor.  We have introduced the notation
\begin{equation}
\stackrel{\sim}{B}_{\mu \nu} = \frac{1}{2} \, \epsilon_{\mu \nu \rho \sigma} \, B^{\rho \sigma} \, ,
\end{equation} 
and our convention for the $ \epsilon$ tensor is $ \epsilon_{0123} =1$.  A brief comment on the completeness of these operators is in order.  $CP$ violating operators containing three $\epsilon$ tensors are reducible to those containing only one such tensor, as the product of two $\epsilon$ tensors can be written as a determinant of metric tensors.  Operators such as $\epsilon_{\mu \nu \rho \sigma}A^{\mu \alpha}B^{\nu}_{\alpha}C^{\rho \beta}D^{\sigma}_{\beta}$, where the $\epsilon$ tensor contracts an index on each field strength tensor, are reducible to those where the $\epsilon$ tensor fully contracts one of the field strength tensors through use of the Schouten identity (see the first reference in~\cite{Trilinear},~\cite{Denner}),
\begin{equation}
g_{\alpha \beta} \epsilon_{\mu \nu \rho \sigma}+g_{\alpha \mu} \epsilon_{\nu \rho \sigma \beta}+g_{\alpha \nu} \epsilon_{\rho \sigma \beta \mu}+g_{\alpha \rho} \epsilon_{\sigma \beta \mu \nu}+g_{\alpha \sigma} \epsilon_{\beta \mu \nu \rho}=0 \, ,
\end{equation}
which states that no tensor antisymmetric in five indices exists in four dimensions.  Similarly, the operators
\begin{eqnarray}
& & B^{\mu \nu}B_{\nu \rho}B^{\rho \sigma} \stackrel{\sim}{B}_{\sigma \mu}, \,\,\,\,
W_{a}^{\mu \nu}W_{a \nu \rho}W_{b}^{\rho \sigma} \stackrel{\sim}{W}_{b \sigma \mu} , \nonumber \\
& & W_{a}^{\mu \nu}W_{a \nu \rho}B^{\rho \sigma} \stackrel{\sim}{B}_{\sigma \mu} , \,\,\,\,
B^{\mu \nu}B_{\nu \rho}W_{a}^{\rho \sigma} \stackrel{\sim}{W}_{a \sigma \mu} \, ,
\end{eqnarray}
can be reduced to those listed in eq. (10) through use of the identity
\begin{equation}
\epsilon_{\mu \alpha \beta \gamma}F^{\nu \alpha}F^{\beta \gamma} = \frac{1}{4} g^{\mu}_{\nu} \epsilon_{\alpha \beta \gamma \delta}F^{\alpha \beta}F^{\gamma \delta} \, .
\end{equation}
This is true for $F^{\mu \nu}$ antisymmetric in four-dimensional spaces (an analog holds in arbitrary dimensions~\cite{Rund}).  Finally, since the processes considered here only involve $W_{3}^{\mu \nu}$, the operators
\begin{equation}
(W_{a \mu \nu}W_{b}^{\mu \nu})(W_{a}^{\rho \sigma} \stackrel{\sim}{W}_{b \rho \sigma} ), \,\,\,\,
W_{a}^{\mu \nu}W_{b \nu \rho}W_{a}^{\rho \sigma} \stackrel{\sim}{W}_{b \sigma \mu}
\end{equation}
become equivalent to the second operator in eq. (10) through use of the above identity, while the operators
\begin{eqnarray}
& &\epsilon_{abc} (W_{a \mu \nu}W_{b}^{\mu \nu})(B^{\rho \sigma}\stackrel{\sim}{W}_{c \rho \sigma} ) , \,\,\,\,
\epsilon_{abc} (B_{\mu \nu}W_{a}^{\mu \nu})(W_{b}^{\rho \sigma} \stackrel{\sim}{W}_{c \rho \sigma} ), \nonumber \\
& & \epsilon_{abc} W_{a}^{\mu \nu}W_{b \nu \rho}W_{c}^{\rho \sigma}\stackrel{\sim}{B}_{\sigma \mu}, \,\,\,\,
\epsilon_{abc}B^{\mu \nu}W_{a \nu \rho}W_{b}^{\rho \sigma} \stackrel{\sim}{W}_{c \sigma \mu} 
\end{eqnarray}
only contribute to processes involving $W^{\pm}$.  We thus see that the seven operators delineated in eq. (10) form a complete and independent set.

Each of these operators will give rise to a number of different $\gamma \gamma \gamma \gamma$, $\gamma \gamma \gamma Z$, and $\gamma \gamma ZZ$ operators, as well as other quartic operators involving three or more $Z$ bosons and various quintic terms.  In this paper we will concentrate on those quartic operators relevant for the scattering processes $\gamma \gamma \rightarrow \gamma \gamma$, $\gamma Z$, and $ZZ$; the other terms cannot be probed in $2 \rightarrow 2$ scattering, and the resulting constraints obtainable on the operators of eq. (10) would be weakened.  Denoting the operators to be studied by ${\cal O}_{i}^{\gamma \gamma}$, ${\cal O}_{i}^{\gamma Z}$, and ${\cal O}_{i}^{ZZ}$, each SU(2) $\times$ U(1) operator will have an expansion of the form
\begin{equation}
{\cal O}_{\alpha}^{SU(2) \times U(1)} = a_{\alpha}^{i}{\cal O}_{i}^{\gamma \gamma} + b_{\alpha}^{i} {\cal O}_{i}^{\gamma Z} + c_{\alpha}^{i} {\cal O}_{i}^{ZZ} + \ldots \,\, ,
\end{equation}
where the $a$, $b$, and $c$ are functions of the weak mixing angle and the ellipsis denotes the neglected quartic and quintic terms.  To determine the independent $\gamma \gamma$, $\gamma Z$, and $ZZ$ structures, we set
\begin{eqnarray}
B^{\mu \nu} &=& {\rm cos}(\theta_W) \, F^{\mu \nu} - {\rm sin}(\theta_W) \, Z^{\mu \nu} \, , \nonumber \\
W_{3}^{\mu \nu} &=& {\rm sin}(\theta_W) \, F^{\mu \nu} + {\rm cos}(\theta_W) \, Z^{\mu \nu} \, ,
\end{eqnarray}
where
\begin{eqnarray}
F_{\mu \nu} &=& \partial_{\mu}A_{\nu} - \partial_{\nu}A_{\mu} \, , \nonumber \\
Z_{\mu \nu} &=& \partial_{\mu}Z_{\nu} - \partial_{\nu}Z_{\mu} \, .
\end{eqnarray}
Doing so, and making use of the identity in eq. (14), leads to the following forms for the ${\cal O}_{i}^{X}$:
\begin{eqnarray}
{\cal O}_{1}^{\gamma \gamma} &:& ( F_{\mu \nu}F^{\mu \nu} ) ( F^{\rho \sigma} \stackrel{\sim}{F}_{\rho \sigma} ) \, , \nonumber \\
{\cal O}_{1-3}^{\gamma Z} &:&  ( F_{\mu \nu}F^{\mu \nu} ) ( F^{\rho \sigma} \stackrel{\sim}{Z}_{\rho \sigma} ) \, , \,\,  ( F_{\mu \nu}Z^{\mu \nu} ) ( F^{\rho \sigma} \stackrel{\sim}{F}_{\rho \sigma} ) \, , \,\,  F^{\mu \nu} F_{\nu \rho} F^{\rho \sigma} \stackrel{\sim}{Z}_{\sigma \mu} \, , \nonumber \\
{\cal O}_{1-5}^{ZZ} &:& ( F_{\mu \nu}F^{\mu \nu} ) ( Z^{\rho \sigma} \stackrel{\sim}{Z}_{\rho \sigma} ) \, , \,\, ( Z_{\mu \nu}Z^{\mu \nu} ) ( F^{\rho \sigma} \stackrel{\sim}{F}_{\rho \sigma} ) \, , \,\, ( F_{\mu \nu}Z^{\mu \nu} ) ( F^{\rho \sigma} \stackrel{\sim}{Z}_{\rho \sigma} ) \, , \nonumber \\  & & F^{\mu \nu} Z_{\nu \rho} F^{\rho \sigma} \stackrel{\sim}{Z}_{\sigma \mu} \, , \,\,  Z^{\mu \nu} F_{\nu \rho} Z^{\rho \sigma} \stackrel{\sim}{F}_{\sigma \mu} \, .
\end{eqnarray}     
We will present the coefficients $a$, $b$, and $c$ when we discuss the relevant process.  

We will organize our study as follows.  A section will be devoted to each of the processes $\gamma \gamma \rightarrow \gamma \gamma$, $\gamma \gamma \rightarrow \gamma Z$, and $\gamma \gamma \rightarrow ZZ$.  In each section we will discuss the properties of the relevant ${\cal O}_{i}^{X}$, focusing on the observable asymmetries associated with each operator and the role of varying the initial polarization states.  We will then discuss the measurement of the various ${\cal O}_{\alpha}^{SU(2) \times U(1)}$, and estimate the sensitivity to these operators that can be obtained at a photon collider.  We will bound each of these operators separately; each is multiplied by an arbitrary coefficient that could cause contributions from different operators to cancel, but we will ignore this possibility.  To preview what we will observe, it will turn out that in certain ${\cal O}_{\alpha}^{SU(2) \times U(1)}$ the structures of eq. (20) will interfere destructively, while in others they will combine constructively, leading to widely varying sensitivity to the ${\cal O}_{\alpha}^{SU(2) \times U(1)}$.

\section{$\gamma \gamma \rightarrow \gamma \gamma$}

In this section we consider the process $ \gamma (k_1) + \gamma(k_2) \rightarrow \gamma(p_1) + \gamma(p_2)$.  We also present and motivate the various parameterizations and approximations used throughout the paper.  A detailed presentation of the Standard Model amplitudes can be found in~\cite{Goun_gg1,Goun_gg2}; we focus here only upon those features relevant to our analysis.

Let us first discuss the properties of the $CP$ violating amplitudes we are considering.  The detailed expressions for these amplitudes, and those for the other processes considered in this paper, can be found in the Appendix.  Bose symmetry implies the relations
\begin{equation}
{\cal M}^{CP}_{abcd} (s,t,u) = {\cal M}^{CP}_{bacd} (s,u,t) = {\cal M}^{CP}_{badc} (s,t,u) \,,
\end{equation}
where $s$, $t$, and $u$ are the usual Mandelstam invariants: $s=(k_1 + k_2)^2$, $t=(k_1 - p_1)^2$, and $u=(k_1 - p_2)^2$.  We have denoted the $CP$ violating amplitudes with the superscript $CP$.  The operators considered in this paper are odd under both $P$ and $T$, which implies
\begin{equation}
{\cal M}^{CP}_{abcd} (s,t,u) = -{\cal M}^{CP}_{-a-b-c-d} (s,t,u) = -{\cal M}^{CP}_{cdab} (s,t,u) \, .
\end{equation}
In addition, the standard crossing symmetries hold.  These relations immediately imply
\begin{equation}
{\cal M}^{CP}_{\pm \pm \pm \pm} = {\cal M}^{CP}_{\pm \mp \pm \mp} = {\cal M}^{CP}_{\pm \mp \mp \pm} =0 \,\, .
\end{equation}
The only non-vanishing amplitudes are $ {\cal M}^{CP}_{\pm \pm \mp \mp}$.  The Standard Model amplitudes are dominated at high energies by the amplitudes ${\cal M}^{SM}_{\pm \pm \pm \pm}$, ${\cal M}^{SM}_{\pm \mp \pm \mp}$, and ${\cal M}^{SM}_{\pm \mp \mp \pm}$.  These amplitudes are primarily imaginary, although the real parts are non-negligible; features of all $\gamma \gamma \rightarrow \gamma \gamma $ amplitudes are discussed in~\cite{Goun_gg1}.  All of the $CP$ odd asymmetries that can be constructed involve the interference of $ {\cal M}^{CP}_{\pm \pm \mp \mp}$ with a Standard Model amplitude; to increase our sensitivity to the anomalous interaction we should attempt to define an observable with interference between $ {\cal M}^{CP}_{\pm \pm \mp \mp}$ and one of the dominant Standard Model terms.  Unfortunately, no observable exists which contains such an interference with one of the large imaginary SM amplitudes.  We can, however, construct the following observable which contains an interference with one of the real pieces of the SM amplitudes:
\begin{equation}
A_{ \gamma \gamma} = \frac{ \int_{0}^{2 \pi} \int_{0}^{2 \pi} \, d \phi_1 \, d \phi_2 \, \left[ \left( \frac{d \sigma}{d \Omega}\right) \delta(\phi_1 - \phi_2 -\pi/4) - \left( \frac{d \sigma}{d \Omega}\right)\delta(\phi_1 - \phi_2 +\pi/4) \right]}{\int_{0}^{2 \pi} \int_{0}^{2 \pi} \, d \phi_1 \, d \phi_2 \, \left[ \left( \frac{d \sigma}{d \Omega}\right) \delta(\phi_1 - \phi_2 -\pi/4) + \left( \frac{d \sigma}{d \Omega}\right)\delta(\phi_1 - \phi_2 +\pi/4) \right]} \, ,
\end{equation}
where $\phi_1$, $\phi_2$ denote the angles of linear polarization introduced in our discussion of the density matrix formalism.  The experimental determination of this asymmetry requires several measurements, as indicated in eq. (24) by the integration over linear polarization directions.  The delta functions in $A_{\gamma \gamma}$ denote the need to maintain a fixed angle between the linear polarization directions of the initial beams.  This observable has been considered previously in~\cite{CP,Choi}.  Referring to eq. (9), and implementing the relations in eqs. (21-22), we can show that the numerator of this asymmetry contains the term
\begin{equation}
A_{\gamma \gamma} \propto \xi_t (x_1) \xi_t (x_2) \, {\rm Re} \left( {\cal M}^{SM}_{++++} \right) {\rm Im} \left( {\cal M}^{CP}_{--++} \right) \, .
\end{equation}
As shown in the Appendix, the $CP$ violating amplitudes are purely imaginary, leading to a non-vanishing result.  As this observable is sensitive only to the degree of linear polarization, we will use the following set of initial parameters:
\begin{equation}
P_{c1} = P_{c2} = 0 \, ; \,\,\,\, P_{t1} = P_{t2} = 1 \, ;
\end{equation}
we note that with no circular polarization the fermion beam polarizations do not enter any expression.  We can simultaneously measure another independent quantity, the denominator of $A_{\gamma \gamma}$.  This is the unpolarized cross section 
\begin{equation}
\frac{d \, \sigma_{unpol}}{d \, {\rm cos}( \theta )} = \frac{1}{256 \pi s} \left( \mid {\cal M}_{++} \mid^2 + \mid {\cal M}_{--} \mid^2 + \mid {\cal M}_{+-} \mid^2 + \mid {\cal M}_{-+} \mid^2 \right) \, ,
\end{equation}
the form of which can also be obtained from eq. (9).

We will now consider the operator relevant to $ \gamma \gamma $ final states in eq. (20).  We write it in the form
\begin{equation}
 {\cal O}_{1}^{\gamma \gamma} = \frac{g^{2}_{eff} e^2}{(\Lambda_{1})^{4}} \, ( F_{\mu \nu}F^{\mu \nu} ) ( F^{\rho \sigma} \stackrel{\sim}{F}_{\rho \sigma} ) \, ,
\end{equation}
where $\Lambda_1$ is the energy scale of the physics leading to these operators, and $g_{eff}$ the associated coupling constant.  We will hereafter set $g_{eff}=1$, effectively absorbing this coupling constant into the definition of $\Lambda_1$.  Although we will quote our sensitivity in terms of the energy scale that can be probed, the reader should realize that this is not exactly the scale associated with the new physics leading to these anomalous couplings.  We have also ignored a possible $\pm 1$ appearing in front; the unpolarized cross section is unaffected by this sign, and $A_{\gamma \gamma}$ only acquires an overall sign change.

In our analysis, we have used the approximate SM amplitudes found in~\cite{Goun_gg2}, valid for $m_{W}^2 / \{s,t,u \} <1 $, which holds for the collider energies considered here.  We will use similar approximate expressions when considering $\gamma \gamma \rightarrow \gamma Z$ and $\gamma \gamma \rightarrow ZZ$.  The features of the SM amplitudes noted earlier occur when the energy of the process becomes greater than several hundred GeV.  In accordance with this approximation we employ the cuts
\begin{equation}
\mid {\rm cos}( \theta ) \mid \leq 0.866, \,\,\, \sqrt{0.4} < x_i < x_{max} \, .
\end{equation}
Throughout this paper, we take the $e^+ e^-$ center of mass energy to be $\sqrt{s}=1000$ GeV when obtaining our search reaches.  In this case, at the endpoints of this region, where $x_1 = x_2 =\sqrt{0.4}$ and ${\rm cos}(\theta) = 0.866$, we have $m_{W}^2 / \mid t \mid \sim 0.24$; this value becomes much smaller as we approach the center of the parameter region, thus validating our approximation.  Although a more detailed analysis taking into account detector properties and experimental cuts would need to use the complete SM expressions, our study captures the salient points.

We show the total cross section and asymmetry versus $\sqrt{s}$ for this operator in fig. (1) for a ``typical'' value of $ \Lambda_{\alpha}$ for purposes of demonstration.  In presenting these results we have assumed an $e^+ e^-$ integrated luminosity $L=500 \, {\rm fb^{-1}}$, which is the quoted yearly integrated luminosity for planned linear colliders~\cite{Tesla}.  However, a determination of $A_{\gamma \gamma}$ requires several separate measurements, and it is certainly not expected that colliders will operate primarily with purely linearly polarized beams, or even primarily in a photon collision mode.  These numbers should be interpreted as results coming from several years of operation and are intended for purposes of illustration.  To provide a more conservative outlook, our quoted sensitivity will be given as a function of integrated luminosity, beginning at the modest value $L=50 \, {\rm fb^{-1}}$.  We have also assumed the value $\alpha =1/137.036$ in this and all other numerical results presented.

At the luminosity $L=500 \, {\rm fb^{-1}}$, the asymmetry is much smaller than the associated errors, and only a change in the total counting rate is statistically observable.  Such an effect can arise from a variety of sources, and the identification of $CP$ violation requires higher luminosities to observe a non-vanishing asymmetry.  

The transcription of these results into statements about the SU(2) $\times$ U(1) operators is simplified by there being only one $\gamma \gamma$ operator.  The coefficients $a^{1}_{SU(2) \times U(1)} $ are 
\begin{eqnarray}
& & a^{1}_{(BB)(BB)} = c^{4}_{W}\, , \,\,\,\, a^{1}_{(WW)(WW)} = s^{4}_{W}\, , \nonumber \\
& & a^{1}_{(BB)(WW)} = a^{1}_{(WW)(BB)} = a^{1}_{(BW)(BW)} = c^{2}_{W} s^{2}_{W} \, ,\nonumber \\
& & a^{1}_{BWBW} = a^{1}_{WBWB} = \frac{1}{4} c^{2}_{W} s^{2}_{W} \, ,
\end{eqnarray}
where $c_{W},s_{W}= {\rm cos}(\theta_{W}),{\rm sin}(\theta_{W})$.  The SU(2) $\times$ U(1) operators of eq. (10) take the form
\begin{equation}
{\cal O }_{\alpha} = \frac{a_{\alpha} e^2}{(\Lambda_{\alpha})^4 } \, ( F_{\mu \nu}F^{\mu \nu} ) ( F^{\rho \sigma} \stackrel{\sim}{F}_{\rho \sigma} ) \, .
\end{equation}
To estimate the value of $\Lambda_{\alpha}$ that can be probed at a $\gamma \gamma$ collider we have performed a combined least-squares fit to the total cross section and asymmetry.  We have assumed standard statistical errors and an additional 1 \% luminosity error in the integrated cross section.  The fit was performed with $\sqrt{s} = 1000$ GeV; approximate results for other energies can be obtained by scaling these numbers.  As can be seen from eq. (30) there are only four different results for this process; these are presented in fig. (2).

Although this asymmetry is not particularly sensitive to $CP$ violation as parameterized by the operators we consider, we note that it is a good test of any $CP$ violating amplitude for which any of the real amplitudes ${\cal M}^{CP}_{\pm \pm \mp \mp}$, ${\cal M}^{CP}_{\pm \mp \pm \mp}$, or ${\cal M}^{CP}_{\pm \mp \mp \pm}$ are non-vanishing.  In these cases terms of the form
\begin{equation}
\xi_t (x_1) \xi_t (x_2) \, {\rm Im} \left( {\cal M}^{SM} \right) {\rm Re} \left( {\cal M}^{CP} \right) 
\end{equation}
appear in the fifth line of eq. (9), leading to a large $A_{\gamma \gamma}$.

\noindent
\begin{figure}[htbp]
\centerline{
\psfig{figure=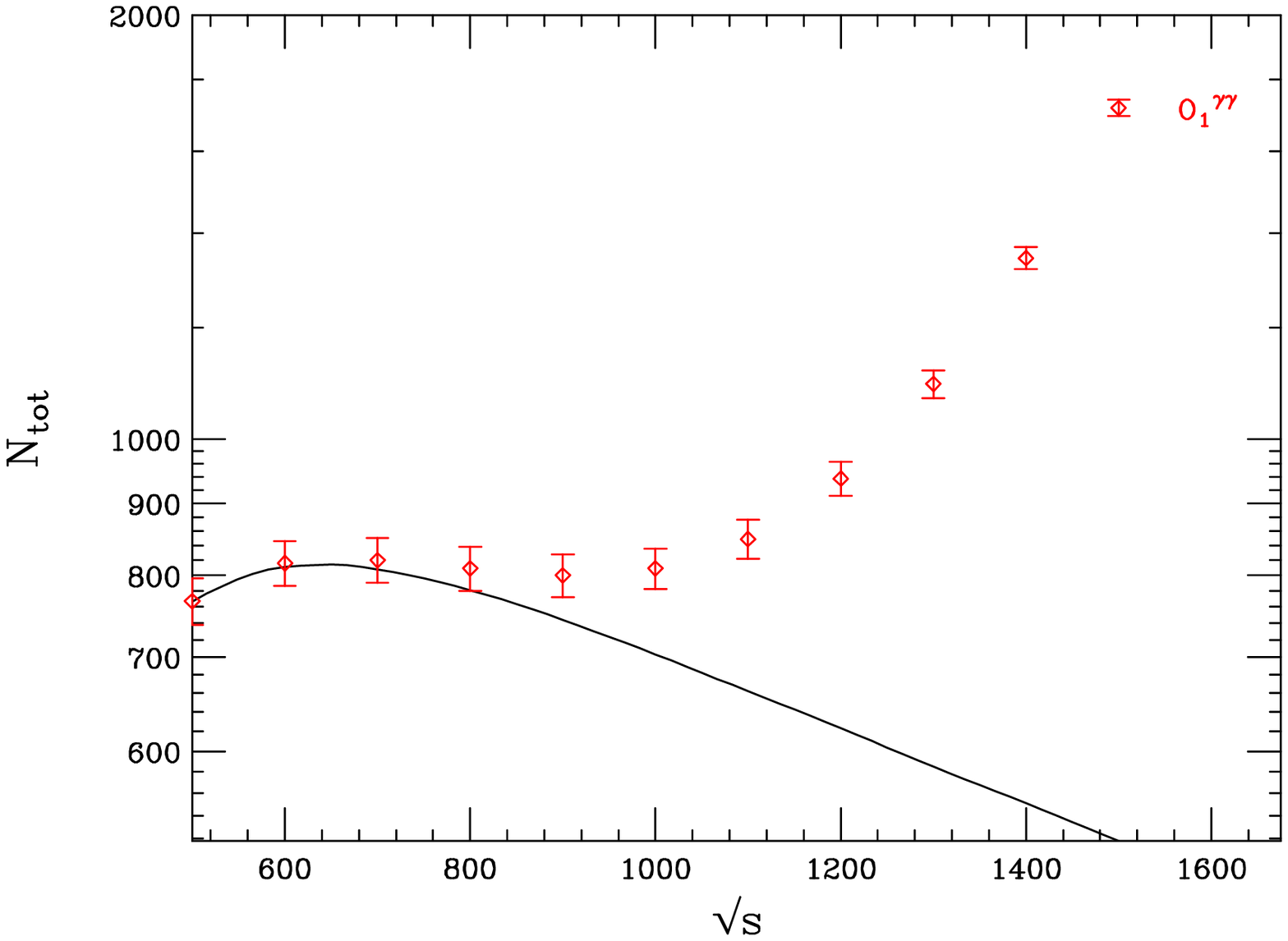,height=8.4cm,width=12.8cm,angle=0}}
\vspace*{1.0cm}
\centerline{
\psfig{figure=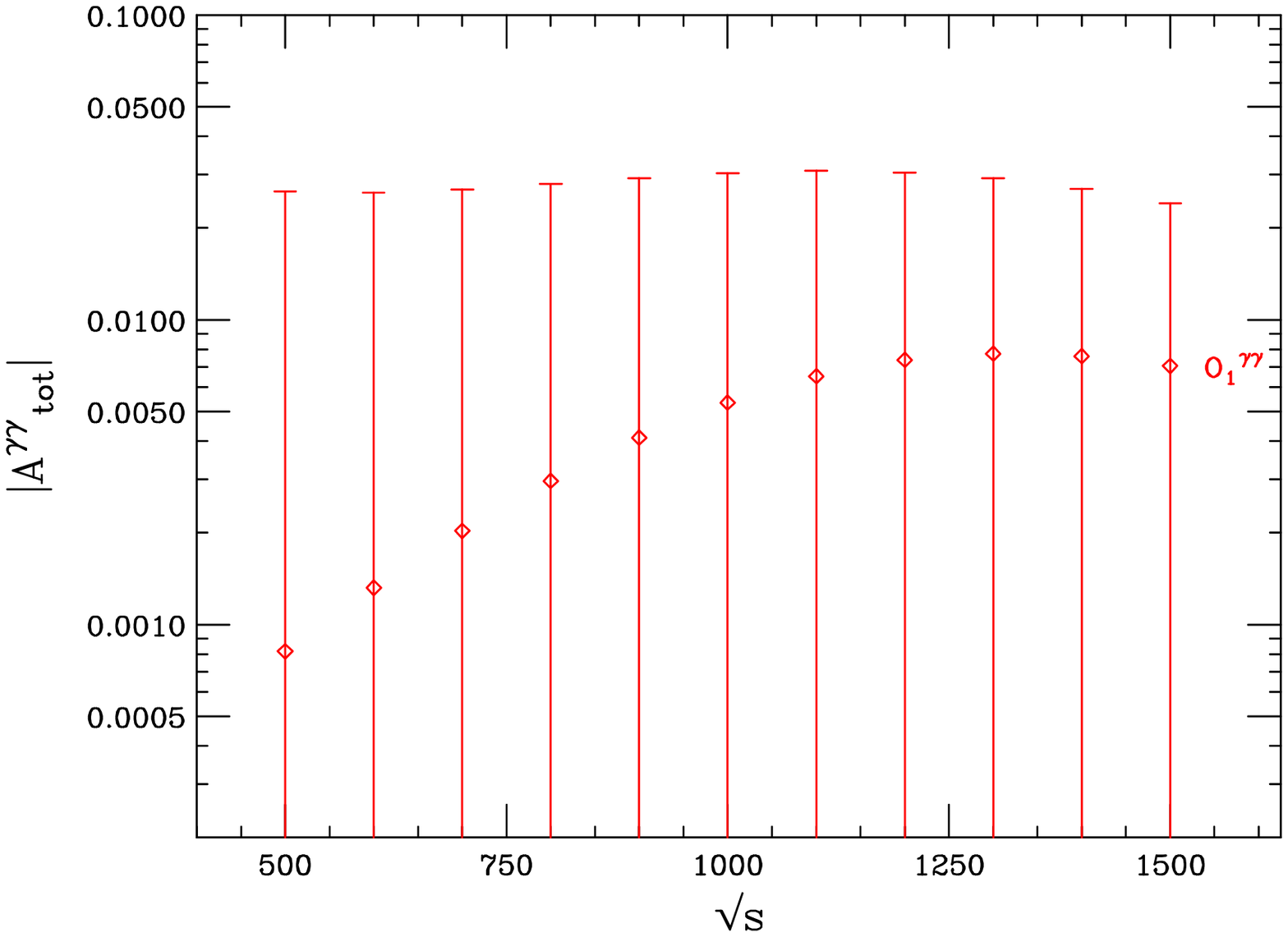,height=8.4cm,width=12.8cm,angle=0}}
\caption{Total cross section (top) and asymmetry (bottom) for the anomalous $ \gamma \gamma$ operator, with $\Lambda_1 = 2 $ TeV and $L=500 \, {\rm fb^{-1}}$.  The bars correspond to the statistical error plus a 1 \% luminosity uncertainty in the event rate; the solid line represents the SM event rate.}
\end{figure}

\noindent
\begin{figure}[htbp]
\centerline{
\psfig{figure=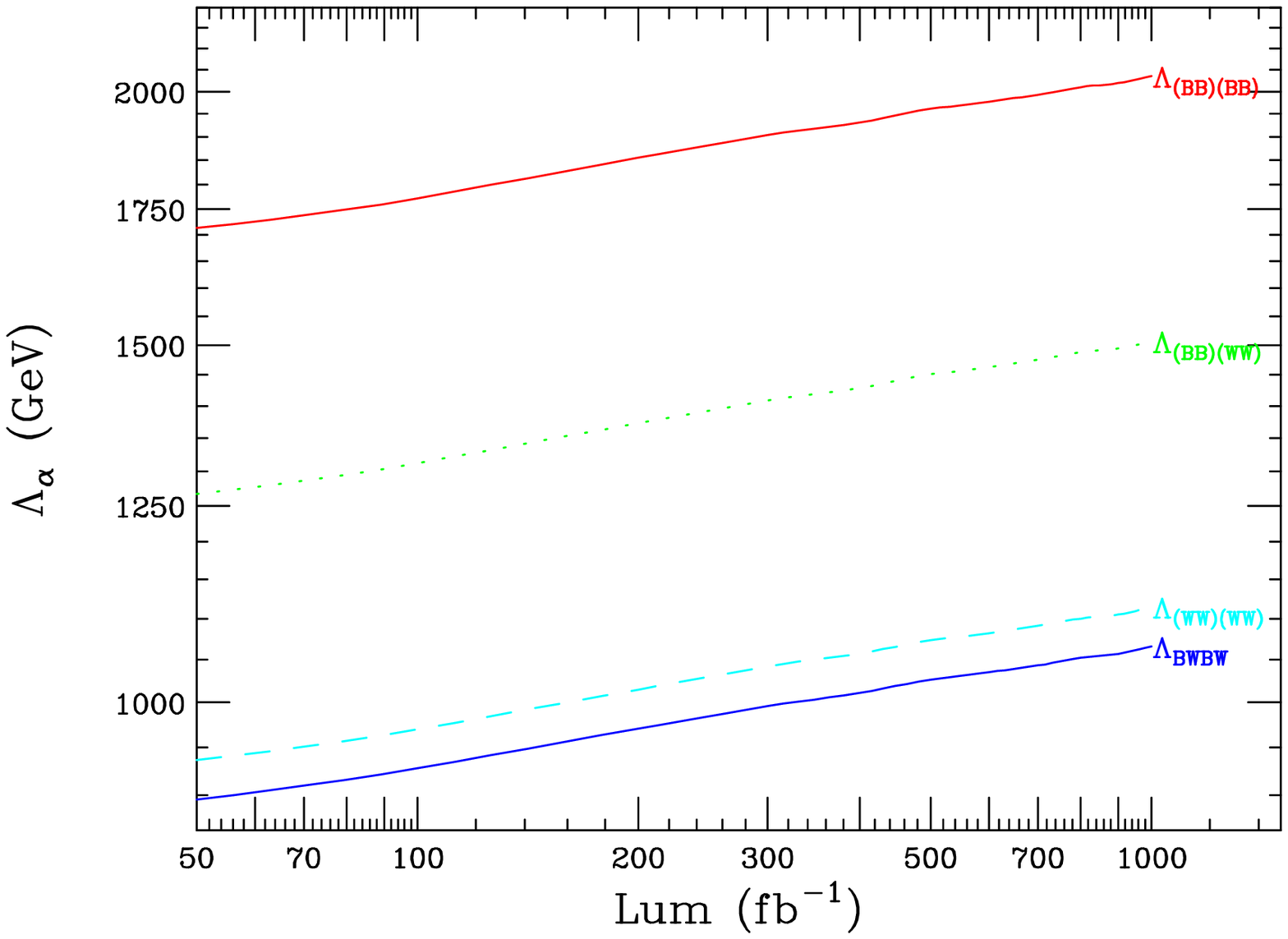,height=9.0cm,width=11.0cm,angle=0}}
\caption{Sensitivity to $\Lambda_{\alpha}$ for each SU(2) $\times$ U(1) operator as a function of integrated luminosity at the 95\% CL.}
\end{figure}

\section{ $ \gamma \gamma \rightarrow \gamma Z$}

In this section we study the effects of anomalous $CP$ violating operators in $\gamma(k_1) + \gamma(k_2) \rightarrow \gamma(p_1) + Z(p_2)$.  A detailed study of the SM amplitudes for this process can be found in~\cite{Goun_gg3}.
  
As in $\gamma \gamma \rightarrow \gamma \gamma$, the $ \gamma Z$ amplitudes satisfy certain relations dictated by Bose symmetry; the anomalous couplings considered satisfy additional relations because of their transformations under $P$.  Let $ \lambda_Z = 0, \pm 1$ denote either longitudinal or transverse Z polarizations.  Bose symmetry implies
\begin{equation}
{\cal M}^{CP}_{abc \lambda_Z} (s,t,u) = (-1)^{1- \lambda_Z}{\cal M}^{CP}_{bac \lambda_Z} (s,u,t) \, ,
\end{equation}
and oddness under $P$ requires
\begin{equation}
{\cal M}^{CP}_{abc \lambda_Z} (s,t,u) = (-1)^{\lambda_Z}{\cal M}^{CP}_{-a-b-c -\lambda_Z} (s,t,u) \, .
\end{equation}
Since the initial and final states are different in this process, the transformation properties under $T$ do not imply any relations between the amplitudes.  The explicit computation of the amplitudes is presented in the Appendix.  Unlike in $\gamma \gamma \rightarrow \gamma \gamma$, there are several surviving amplitudes.

The structure of the SM amplitudes for $ \gamma \gamma \rightarrow \gamma Z$ is similar to that for $\gamma \gamma \rightarrow \gamma \gamma$, and even more pronounced.  At high energies the process is dominated by the imaginary parts of the amplitudes ${\cal M}^{SM}_{\pm \pm \pm \pm}$, ${\cal M}^{SM}_{\pm \mp \pm \mp}$, and ${\cal M}^{SM}_{\pm \mp \mp \pm}$; all other amplitudes are completely negligible~\cite{Goun_gg3}.  As before, we must construct an observable that contains an interference of a $CP$ odd amplitude with one of the dominant SM terms.  The asymmetry considered in the previous section contains only interference with the real parts of the SM amplitudes, and is therefore unacceptable.  For interference effects between imaginary $CP$ odd and $CP$ even amplitudes, we must consider the following asymmetry which is measurable with circularly polarized beams:
\begin{equation}
A_{\gamma Z} = \frac{\left( \frac{d \sigma}{d \Omega} \right)_+ - \left( \frac{d \sigma}{d \Omega} \right)_- }{\left( \frac{d \sigma}{d \Omega} \right)_+ + \left( \frac{d \sigma}{d \Omega} \right)_- } \,\, .
\end{equation}
This asymmetry has been considered previously in~\cite{CP}.  The subscripts $\pm$ denote the initial polarization states of the laser and fermion beams, which we will now discuss.  Setting $ P_{t1} = P_{t2} =0$, we are left with four parameters describing the inital state polarization: $P_{e1}$, $P_{e2}$, $P_{l1}$, and $P_{l2}$.  We will set $|P_e |=0.9$ and $|P_l | =1.0$, consistent with the expected capabilities of future facilities~\cite{Tesla}, and label our inital states as $(P_{e1},P_{l1},P_{e2},P_{l2})$.  In the SM, there are six independent states: $(++++), \, (+++-), \, (++--), \, (+-+-), \, (-++-),$ and $(+---)$, where, for example, $(+-+-)$ means $P_{e1}=0.9$, $P_{l1}=-1.0$, $ P_{e2}=0.9$, and $P_{l2}=-1.0$.  States obtained by an overall sign flip are identical; for example, $(+-+-)$ and $(-+-+)$ lead to the same observables.  This is not true when $CP$ violating interactions are present.  The asymmetry of eq. (35), where the subscript $+$ refers to a given inital state and $-$ to the state obtained by flipping the signs of the polarizations, will vanish in the SM and be non-zero in the presence of the anomalous couplings.  In addition to this asymmetry we can simultaneously measure another independent quantity, the denominator of the asymmetry.  Referring to the ensemble average in eq. (9) and using the relations in eqs. (33-34), we see that this denominator is just twice the polarized cross section,
\begin{eqnarray}
\frac{d \sigma_{pol}}{d {\rm cos}(\theta)} &=& \frac{1}{128 \pi s} \bigg[ \left(1+ \xi_{c}(x_1) \xi_{c}(x_2) \right) \left(  \mid {\cal M}_{++} \mid^2 + \mid {\cal M}_{--} \mid^2 \right) \nonumber \\ & & + \left(1- \xi_{c}(x_1) \xi_{c}(x_2) \right) \left(  \mid {\cal M}_{+-} \mid^2 + \mid {\cal M}_{-+} \mid^2 \right) \bigg] \, .
\end{eqnarray}

The operators relevant to $ \gamma Z$ final states can be found in eq. (20); we will write them as follows:
\begin{eqnarray}
 {\cal O}_{1}^{\gamma Z} &=& \frac{e^2}{(\Lambda_{1})^4} \, ( F_{\mu \nu}F^{\mu \nu} ) ( F^{\rho \sigma} \stackrel{\sim}{Z}_{\rho \sigma} ) \nonumber \\
 {\cal O}_{2}^{\gamma Z} &=& \frac{e^2}{(\Lambda_{2})^4} \, ( F_{\mu \nu}Z^{\mu \nu} ) ( F^{\rho \sigma} \stackrel{\sim}{F}_{\rho \sigma} ) \nonumber \\
{\cal O}_{3}^{\gamma Z} &=& \frac{e^2}{(\Lambda_{3})^4} \, F^{\mu \nu} F_{\nu \rho} F^{\rho \sigma} \stackrel{\sim}{Z}_{\sigma \mu} \, .
\end{eqnarray}
Our philosophy in parameterizing these operators is identical to that discussed in the $\gamma \gamma \rightarrow \gamma \gamma$ section; there is again a coupling constant $g_{eff}$ associated with each operator that we have set equal to 1, and a possible factor of $\pm1$ appearing in each operator that we have ignored.  The detailed amplitudes are presented in the Appendix; here we will discuss several properties relevant to our analysis.  The first is that the amplitudes for longitudinally polarized $Z$ states are suppressed relative to those with transverse $Z$ states by a factor of $m_{Z} / \sqrt{s}$, as dictated by the Goldstone boson equivalence theorem.  This, and the fact that the tagging efficiency for longitudinal states is at best 10\% , means that we will not explore the effect of selecting final state polarizations in our analysis.  

Before presenting our results, let us examine $A_{\gamma Z}$ in more detail.  Our discussion will illustrate the complementary information  that can be obtained from examining all of the possible initial polarizations.  Concentrating on the numerator of $A_{\gamma Z}$, we see that 
\begin{eqnarray}
A_{\gamma Z} & \propto & \left (\xi_{c}(x_{1}) + \xi_{c}(x_{2}) \right) \left[ \mid {\cal M}_{++} \mid^2 - \mid {\cal M}_{--} \mid^2 \right] \nonumber \\ & & + \left (\xi_{c}(x_{1}) - \xi_{c}(x_{2}) \right) \left[ \mid {\cal M}_{+-} \mid^2 - \mid {\cal M}_{-+} \mid^2 \right] \,\, ,
\end{eqnarray}
where a sum over final state helicities is implied.  The first term is symmetric under the interchange of the two initial photons, which implies symmetry under the transformation ${\rm cos}(\theta) \leftrightarrow -{\rm cos}(\theta)$ and under the interchange of the initial polarization states, $(P_{e1},P_{l1}) \leftrightarrow (P_{e2},P_{l2})$.  The second term is antisymmetric under both of these exchanges.  By selecting an appropriate initial polarization state, we can isolate each term.  For example, the 
choice $(++++)$ is symmetric under the interchange of the two initial polarizations; with this choice only the first term of $A_{\gamma Z}$ will contribute, whereas the choice $(-++-)$ is antisymmetric under interchange of the two initial polarizations, and with this selection only the second term contributes.  A choice such as $(+---)$ is of mixed symmetry, and is sensitive to both terms in $A_{\gamma Z}$.  Such a variation of the inital states allows us to isolate the various anomalous amplitudes which contribute to $A_{\gamma Z}$ is sensitive; with symmetric initial polarizations it is sensitive to ${\cal M}^{CP}_{\pm \pm \pm \pm}$, while with antisymmetric polarizations it is sensitive to ${\cal M}^{CP}_{\pm \mp \pm \mp}$ and ${\cal M}^{CP}_{\pm \mp \mp \pm}$.

The four largest asymmetries for ${\cal O}_{1}$ are presented in fig. (3) as a function of ${\rm cos}(\theta)$, employing the same cuts as used in $\gamma \gamma \rightarrow \gamma \gamma$.  At $L=500 \, {\rm fb}^{-1}$ and $\Lambda_{1} = 2$ TeV, the ``symmetric'' asymmetries are statistically significant throughout the entire angular region, while the ``antisymmetric'' asymmetries are significant in the outer regions.  In fig. (4) we present the total cross section versus $\sqrt{s}$ for each inital polarization state, and the integrated asymmetry versus $\sqrt{s}$ for the four symmetric initial states.  Although a deviation in the cross section is not seen in any polarization state until high energies, the integrated asymmetry becomes quite large at low energies, suggesting that it provides a sensitive test of the anomalous couplings under consideration.  So far we have only considered the operator ${\cal O}_{1}^{\gamma Z}$; the distributions for the other two operators differ only in relative sign and overall magnitude at high energies.  The asymmetries for polarization states $(+-+-)$ and $(-++-)$ for all three operators are shown in fig. (5); those for ${\cal O}_{1}^{\gamma Z}$ and ${\cal O}_{2}^{\gamma Z}$ are of similar magnitude but opposite sign.  This can be 
understood in a simple way.  At high energies, a transversely polarized $Z$ effectively looks like a photon.  When computing the amplitudes for the various operators in eq. (37), we must include all possible permutations of the three photons, as they are identical particles (for example, denoting the field strength tensors of the three photons by 1,2, and 3, the six permutations required to compute ${\cal O}_{1}^{\gamma Z}$ become $(12)(3Z)$, $(21)(3Z)$, $(13)(2Z)$, $(31)(2Z)$, $(23)(1Z)$, and $(32)(1Z)$).  The permutations required for ${\cal O}_{1}^{\gamma Z}$ and ${\cal O}_{2}^{\gamma Z}$ are
 identical to the permutations necessary to compute the amplitudes for the four photon operator of eq. (28); we can verify using the explicit forms of the amplitudes in the Appendix that the sum of ${\cal M}_{1}^{\gamma Z}(\lambda_1,\lambda_2;\lambda_3,\lambda_4)$ and ${\cal M}_{2}^{\gamma Z}(\lambda_1,\lambda_2;\lambda_3,\lambda_4)$ gives ${\cal M}_{1}^{\gamma \gamma}(\lambda_1,\lambda_2;\lambda_3,\lambda_4)$, up to an overall constant.  Since $T$ violation leads to vanishing of the amplitudes  ${\cal M}^{CP}_{\pm \pm \pm \pm}$, ${\cal M}^{CP}_{\pm \mp \pm \mp}$ and ${\cal M}^{CP}_{\pm \mp \mp \pm}$ in $\gamma \gamma \rightarrow \gamma \gamma$, the asymmetries for ${\cal O}_{1}^{\gamma Z}$ and ${\cal O}_{2}^{\gamma Z}$ must therefore be of opposite sign.  SU(2) $ \times$ U(1) operators whose coefficients $b^{1}_{\alpha}$, $b^{2}_{\alpha}$ are of the same sign will see destructive interference between  ${\cal O}_{1}^{\gamma Z}$ and ${\cal O}_{2}^{\gamma Z}$, while those whose coefficients have opposite signs will see constructive interference.  We will see similar behavior when investigating $\gamma \gamma \rightarrow ZZ$.

Table (1) presents the coefficients $b^{i}_{\alpha}$ relating the ${\cal O}^{SU(2) \times U(1)}_{\alpha}$ to the ${\cal O}_{i}^{\gamma Z}$.
\begin{table}
\centering
\begin{tabular}{|l|l|l|l|} \hline\hline
$b^{i}_{\alpha}$ & $ {\cal O}_{1}^{\gamma Z}$ & $ {\cal O}_{2}^{\gamma Z}$ &   ${\cal O}_{3}^{\gamma Z}$ \\ \hline ${\cal O}_{(BB)(BB)}$ & $-2s_{W}c^{3}_{W}$ & $-2s_{W}c^{3}_{W}$ & $0$ \\  ${\cal O}_{(WW)(WW)}$ & $2c_{W}s^{3}_{W}$ & $ 2c_{W}s^{3}_{W}$ & $0$ \\  ${\cal O}_{(BB)(WW)}$ & $2s_{W}c^{3}_{W}$ & $-2c_{W}s^{3}_{W}$ & $0$ \\  ${\cal O}_{(WW)(BB)}$ & $-2c_{W}s^{3}_{W}$ & $2s_{W}c^{3}_{W}$ & $0$ \\  ${\cal O}_{(BW)(BW)}$ & $c_{W}s_{W} (c^{2}_{W} - s^{2}_{W})$ & $c_{W}s_{W} (c^{2}_{W} - s^{2}_{W})$ & $0$ \\  ${\cal O}_{BWBW}$ & $0$ & $\frac{1}{4} (s_{W}c^{3}_{W} - 2 c_{W}s^{3}_{W} )$ & $s_{W}c^{3}_{W}$ \\  ${\cal O}_{WBWB}$ & $0$ & $-\frac{1}{4} (c_{W}s^{3}_{W} - 2 s_{W}c^{3}_{W} )$ & $-c_{W}s^{3}_{W}$ \\ \hline\hline
\end{tabular}
\caption{Coefficients $b^{i}_{\alpha}$ relating the ${\cal O}_{\alpha}^{SU(2) \times U(1)}$ to the ${\cal O}_{i}^{\gamma Z}$. }
\end{table}
The $\gamma Z$ operators will interfere destructively in ${\cal O}_{(BB)(BB)}$,  ${\cal O}_{(WW)(WW)}$, and ${\cal O}_{(BW)(BW)}$ and constructively in ${\cal O}_{(BB)(WW)}$ and ${\cal O}_{(WW)(BB)}$.  To estimate the value of $\Lambda_{\alpha} $ that can be probed at a $\gamma \gamma $ collider we have performed a combined least-squares fit to the normalized binned cross section, binned asymmetry, and total cross section, with $ \sqrt{s}=1000$ GeV for the two polarization states $(+-+-)$ and $(-++-)$.  These two choices are chosen to illustrate the sensitivities obtainable from both symmetric and asymmetric initial polarizations; the search reaches from the remaining 
polarization states are similar.  The results are presented in fig. (6).  Here, we have included only five of the SU(2) $\times$ U(1) operators; ${\cal O}_{(BB)(WW)}$ and ${\cal O}_{(WW)(BB)}$ differ only in the sign of their asymmetries, as do ${\cal O}_{BWBW}$ and ${\cal O}_{WBWB}$, and hence have identical discovery regions.  Remembering that our operators are of dimension eight, and that the anomalous amplitudes therefore scale as ${\cal M} \sim s^2 / \Lambda^4 $, we see that the process $\gamma \gamma \rightarrow \gamma Z$ is quite sensitive to the $CP$ violating operators whose asymmetries interfere constructively.  The attractiveness of this process is enhanced by the ease with which it can be experimentally reconstructed.

In summary, we find that this process is a sensitive probe of $CP$ violation in the gauge boson sector.  The asymmetry considered here would be useful for any anomalous amplitudes ${\cal M}^{CP}_{\pm \pm \pm \pm}$, ${\cal M}^{CP}_{\pm \mp \pm \mp}$ and ${\cal M}^{CP}_{\pm \mp \mp \pm}$ which contain imaginary parts.

\noindent
\begin{figure}[htbp]
\centerline{
\psfig{figure=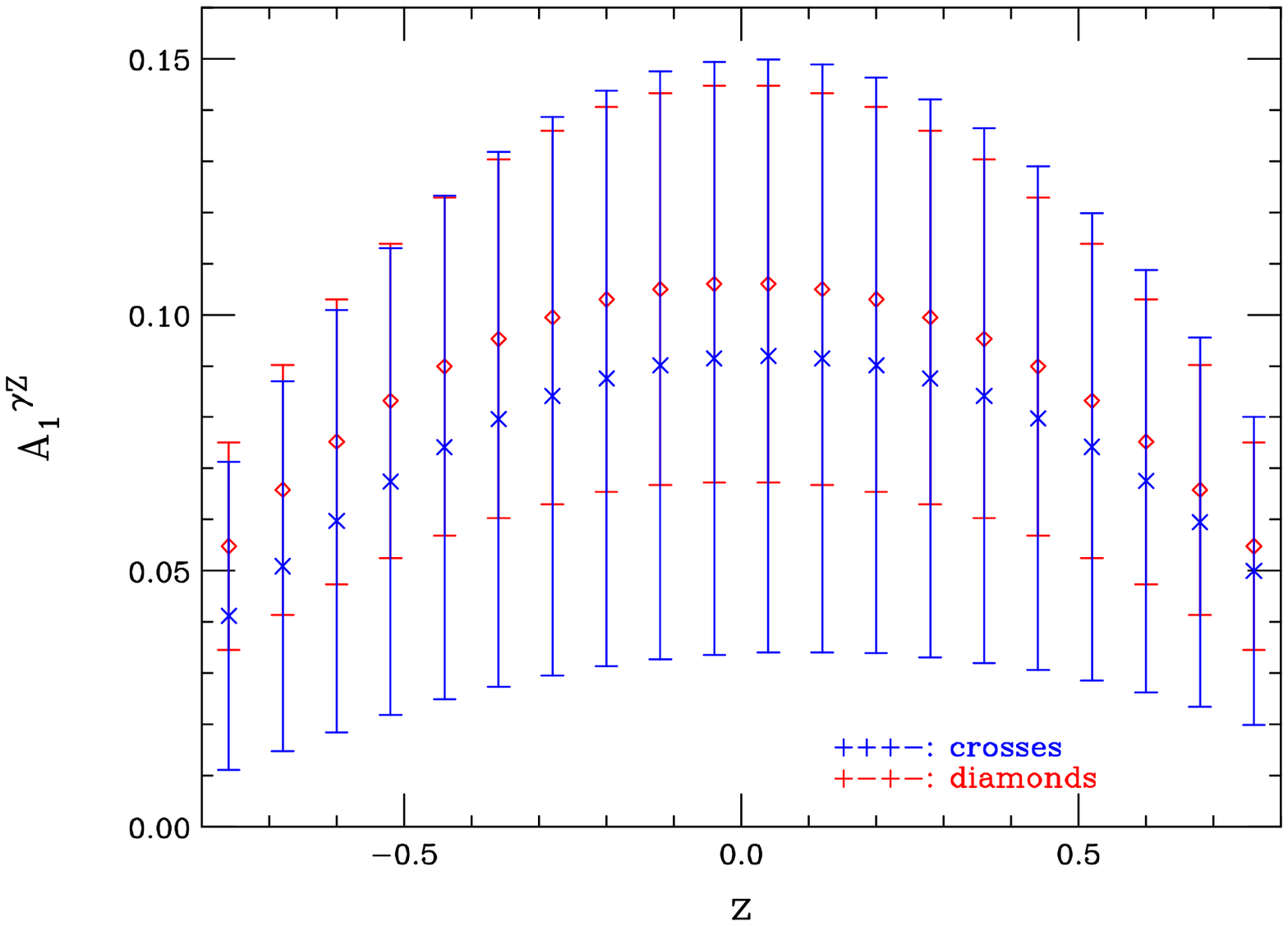,height=8.4cm,width=12.8cm,angle=0}}
\vspace*{1.0cm}
\centerline{
\psfig{figure=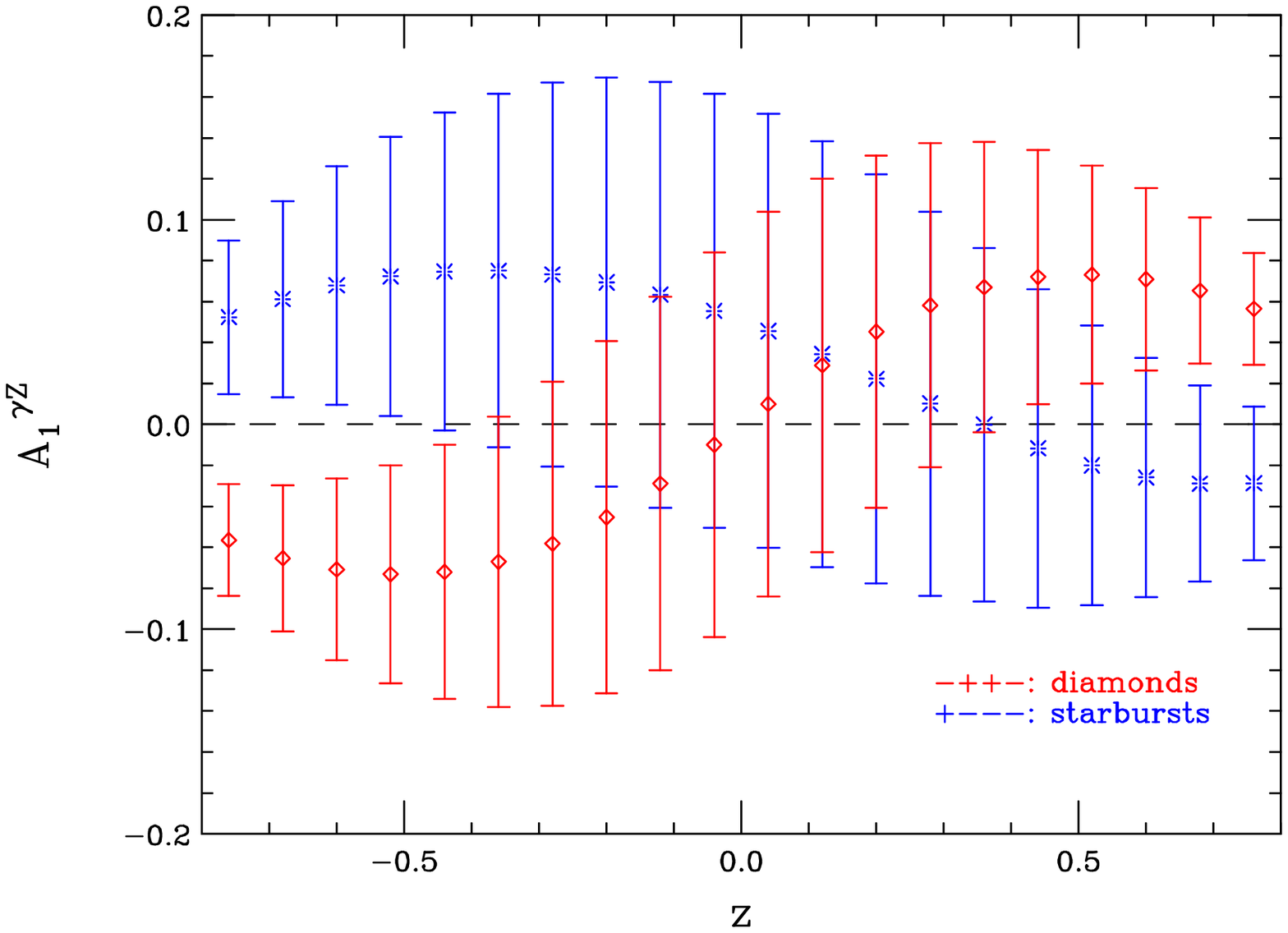,height=8.4cm,width=12.8cm,angle=0}}
\caption{``Primarily symmetric'' asymmetries (top) and ``primarily antisymmetric'' asymmetries (bottom) for ${\cal O}_{1}^{\gamma Z}$, with $ \Lambda_1 = 2$ TeV, $L=500 \, {\rm fb^{-1}}$, and $\sqrt{s}=1000$ GeV.  The bars indicate the corresponding statistical error. }
\end{figure}

\noindent
\begin{figure}[htbp]
\centerline{
\psfig{figure=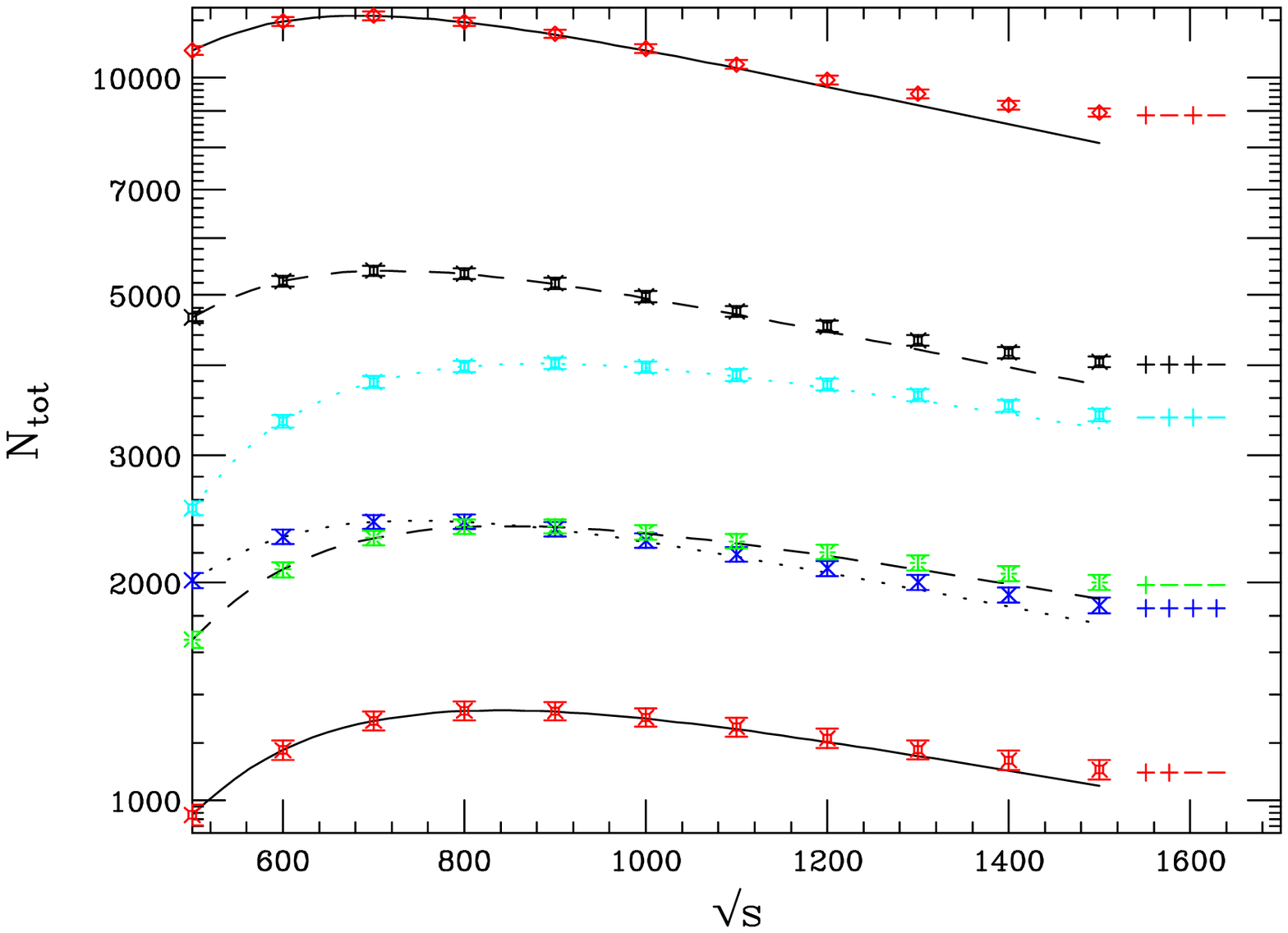,height=8.4cm,width=12.8cm,angle=0}}
\vspace*{1.0cm}
\centerline{
\psfig{figure=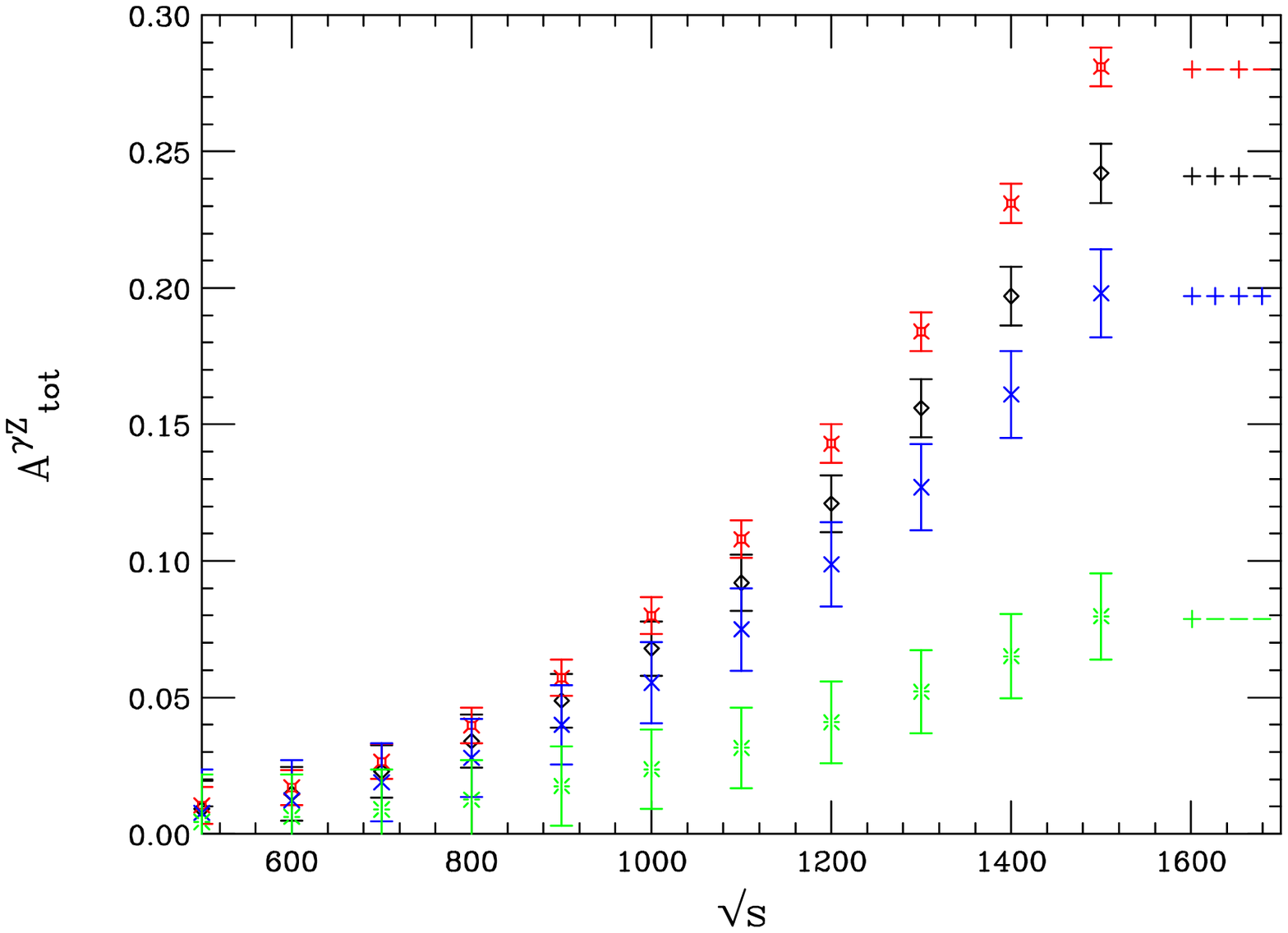,height=8.4cm,width=12.8cm,angle=0}}
\caption{Total (polarized) counting rate (top) and integrated asymmetry (bottom) versus $\sqrt{s}$ for ${\cal O}_{1}^{\gamma Z}$, with $\Lambda_1 = 2$ TeV and $L=500 \, {\rm fb^{-1}}$.  The bars indicate the corresponding statistical error, as well as a 1 \% luminosity uncertainty in the event rate.  The solid curves represent the SM event rates.}
\end{figure}

\noindent
\begin{figure}[htbp]
\centerline{
\psfig{figure=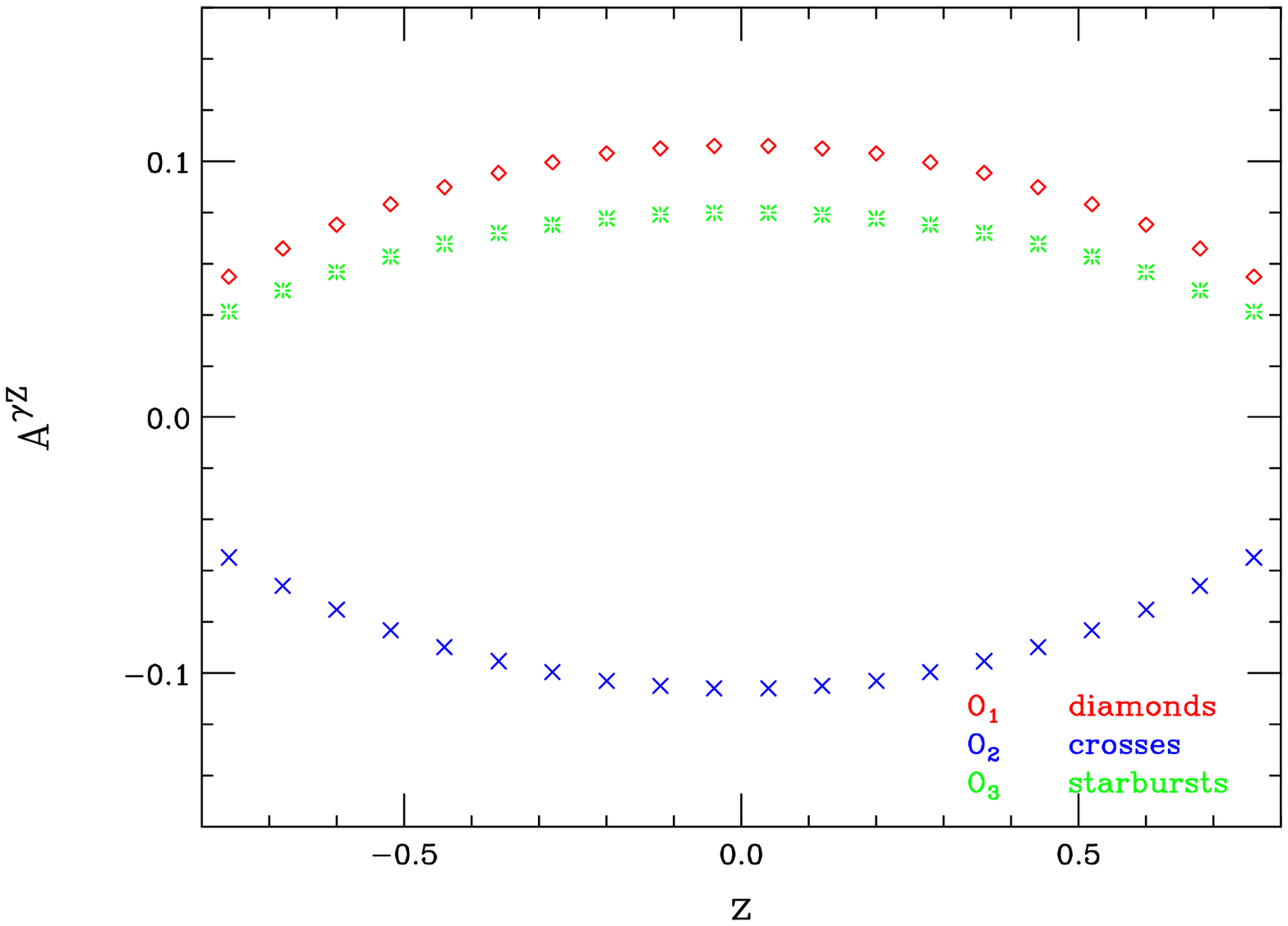,height=8.4cm,width=12.8cm,angle=0}}
\vspace*{1.0cm}
\centerline{
\psfig{figure=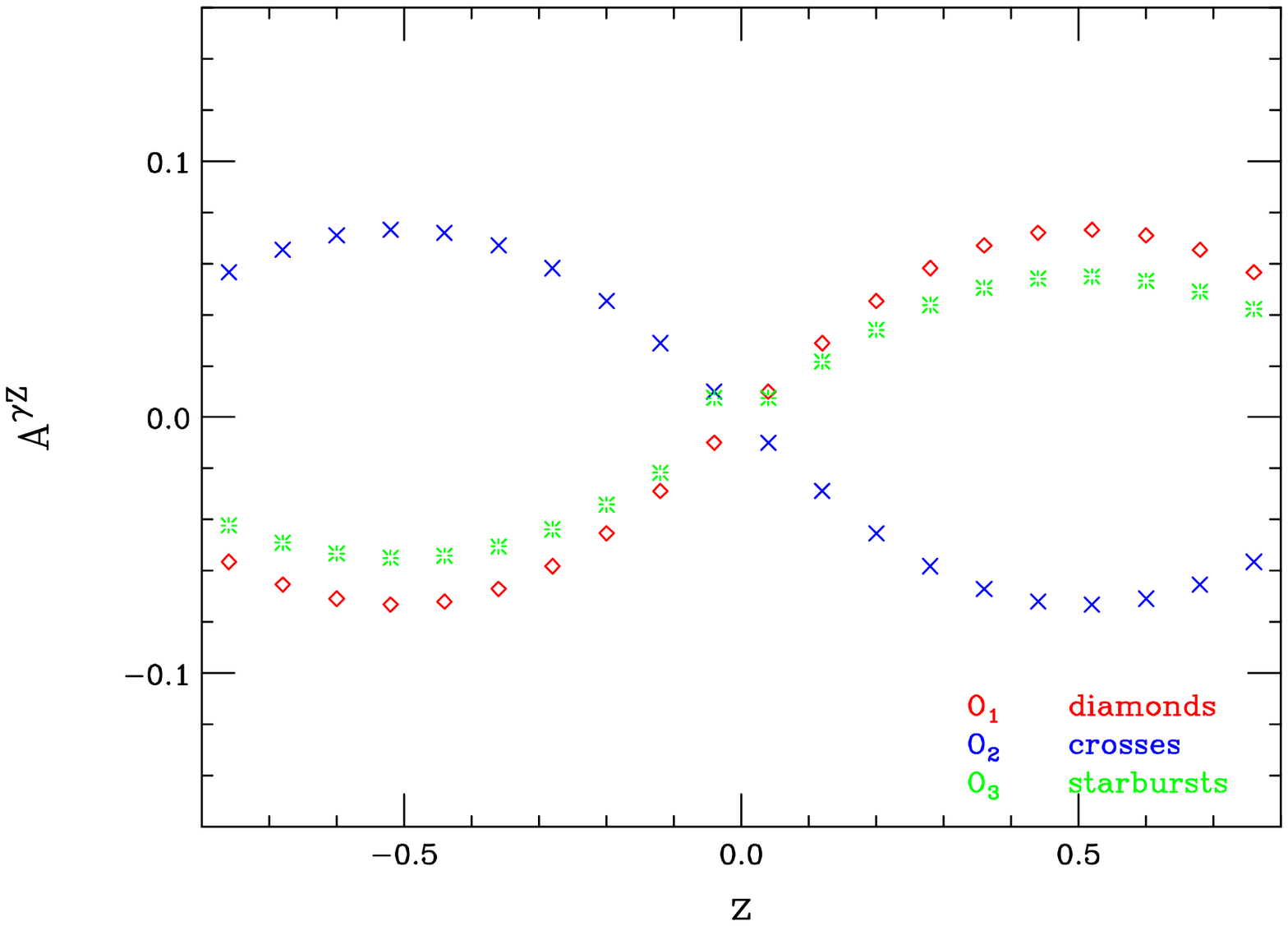,height=8.4cm,width=12.8cm,angle=0}}
\caption{Comparison of $A_{\gamma Z}$ for the polarization states $(+-+-)$ (top) and $(-++-)$ (bottom), with $\Lambda_i = 2$ TeV and $L=500 \, {\rm fb^{-1}}$.  Here, the statistical error bars have been suppressed.}
\end{figure}

\noindent
\begin{figure}[htbp]
\centerline{
\psfig{figure=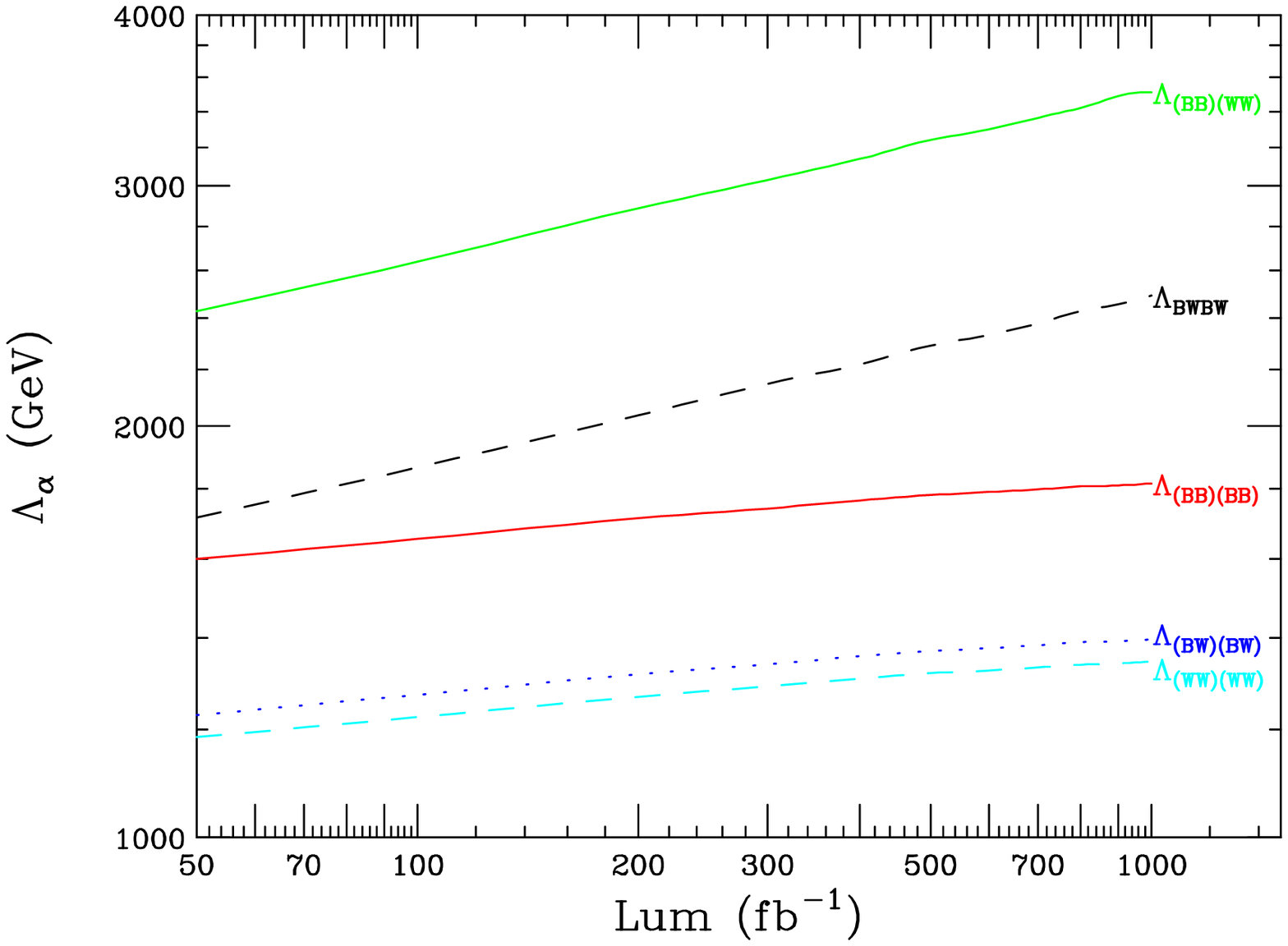,height=8.4cm,width=12.8cm,angle=0}}
\vspace*{1.0cm}
\centerline{
\psfig{figure=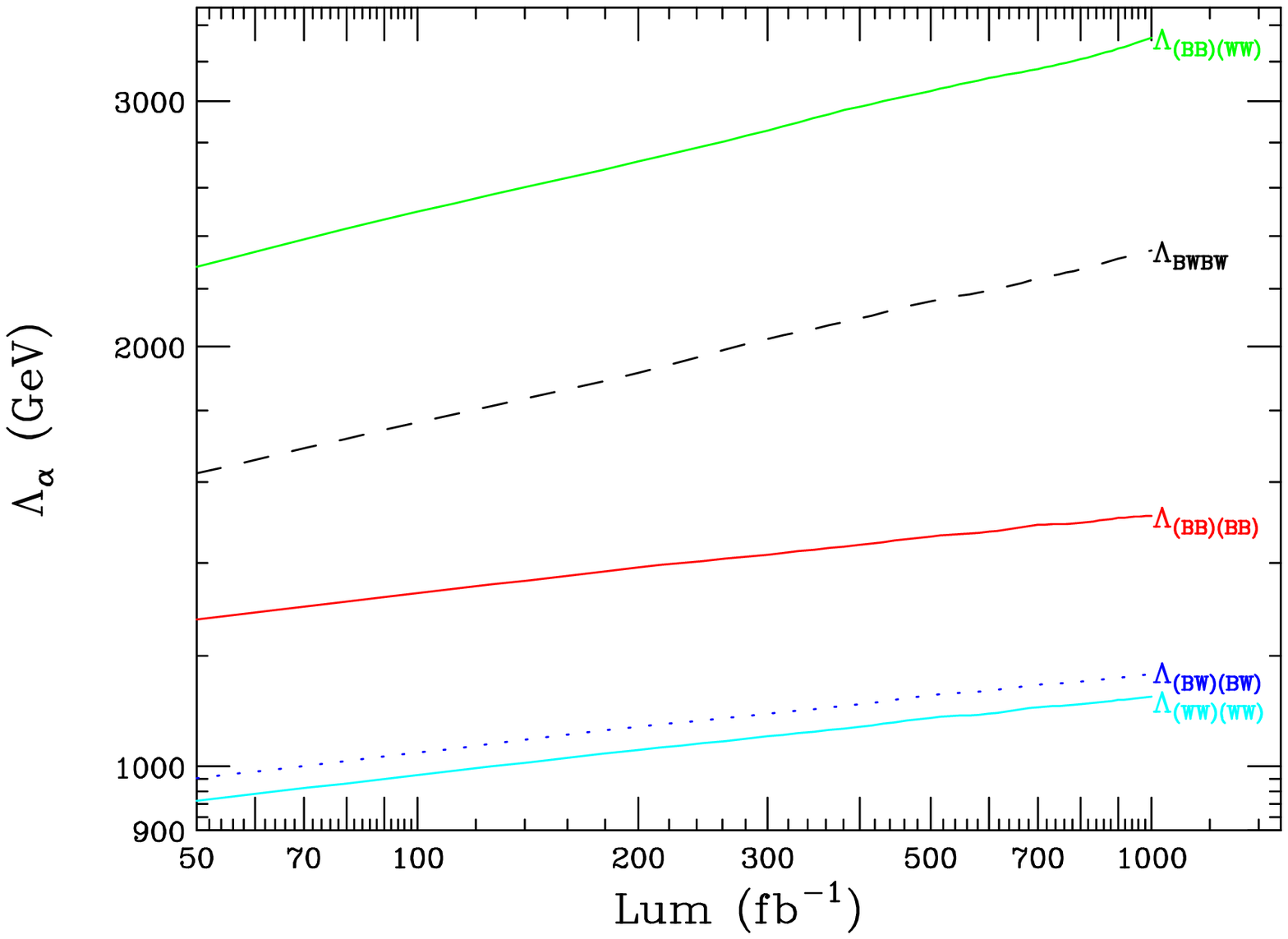,height=8.4cm,width=12.8cm,angle=0}}
\caption{Search reach for $\Lambda_{\alpha}^{\gamma Z}$ for polarization state $(+-+-)$ (top), and $(-++-)$ (bottom).  All of the quoted sensitivities are at the 95 \% CL.}
\end{figure}

\section{ $ \gamma \gamma \rightarrow Z Z$ }

In this section we will discuss the measurement of $CP$ violation in $\gamma(k_1) + \gamma(k_2) \rightarrow Z(p_1) + Z(p_2)$.  A detailed description of the SM process is given in~\cite{Goun_gg4}.
 
We first briefly review the constraints imposed upon the $ ZZ$ amplitudes by Bose symmetry and parity violation.  Let $ \lambda_{i} = 0, \pm 1 $ denote the $Z$ polarizations.  Bose symmetry implies
\begin{equation}
{\cal M}^{CP}_{ab \lambda_1 \lambda_2}(s,t,u) = (-1)^{ \lambda_{1} - \lambda_{2}} {\cal M}^{CP}_{ba \lambda_{1} \lambda_{2}}(s,u,t) = {\cal M}^{CP}_{ab \lambda_2 \lambda_1}(s,u,t) \,\, , 
\end{equation}
while oddness under $P$ implies
\begin{equation}
{\cal M}^{CP}_{ab \lambda_1 \lambda_2}(s,t,u) = (-1)^{ \lambda_{1} - \lambda_{2}-1} {\cal M}^{CP}_{-a -b - \lambda_1 - \lambda_2}(s,t,u) \,\, .
\end{equation}
These relations lead to the vanishing of the amplitudes $ { \cal M}^{CP}_{\pm \mp \pm \mp}$ and ${\cal M}^{CP}_{\pm \mp \mp \pm}$.  However, unlike in $\gamma \gamma \rightarrow \gamma \gamma$, the amplitude ${ \cal M}^{CP}_{\pm \pm \pm \pm}$ is non-zero.  Explicit formulae for these amplitudes are given in the Appendix.

The SM amplitudes for this process are similar to those for $\gamma \gamma \rightarrow \gamma \gamma$ and $\gamma \gamma \rightarrow \gamma Z$ in that they are completely dominated at high energies by the imaginary parts of ${\cal M}^{SM}_{\pm \pm \pm \pm}$, ${\cal M}^{SM}_{\pm \mp \pm \mp}$, and ${\cal M}^{SM}_{\pm \mp \mp \pm}$.  However, they differ in one important aspect.  At one loop they receive contributions from the 
Higgs sector through the one loop $h \gamma \gamma$ vertex, and are therefore not independent of the mechanism of electroweak symmetry breaking.  The contributions to this process from a neutral Higgs, such as the one present in the SM, have been given in~\cite{Goun_gg4}; for a Higgs mass $m_{h} < 2m_{Z}$ (which includes the value $ m_h \cong 115$ GeV suggested by LEP), this contribution is negligible.  For larger values of the Higgs mass, the two $Z$ bosons will reconstruct to $m_h$ for this contribution, making it separable from continuum $ZZ$ production.  We thus neglect the Higgs contribution in our analysis.

The structure of both the SM and the anomalous amplitudes are similar to those in $\gamma \gamma \rightarrow \gamma Z$; in particular, the relevant asymmetry that contains interference between the large imaginary SM amplitudes and the anomalous amplitudes is again given by
\begin{equation}
A_{Z Z} = \frac{\left( \frac{d \sigma}{d \Omega} \right)_+ - \left( \frac{d \sigma}{d \Omega} \right)_- }{\left( \frac{d \sigma}{d \Omega} \right)_+ + \left( \frac{d \sigma}{d \Omega} \right)_- } \,\, .
\end{equation}
Varying the intial polarization state $(P_{e1},P_{l1},P_{e2},P_{l2})$ yields less information than in $\gamma \gamma \rightarrow \gamma Z$, however, for reasons discussed below.  The numerator of the asymmetry again takes the form
\begin{eqnarray}
A_{Z Z} & \propto & \left (\xi_{c}(x_{1}) + \xi_{c}(x_{2}) \right) \left[ \mid {\cal M}_{++} \mid^2 - \mid {\cal M}_{--} \mid^2 \right] \nonumber \\ & & + \left (\xi_{c}(x_{1}) - \xi_{c}(x_{2}) \right) \left[ \mid {\cal M}_{+-} \mid^2 - \mid {\cal M}_{-+} \mid^2 \right] \,\, ,
\end{eqnarray}
but the vanishing of $ { \cal M}^{CP}_{\pm \mp \pm \mp}$ and $ {\cal M}^{CP}_{\pm \mp \mp \pm}$ implies that the second term in eq. (42) equals zero.  We will only observe the ``symmetric'' asymmetries, and the ``antisymmetric'' asymmetries for the initial states $(-++-)$ and $(++--)$ will vanish.  We can again simultaneously measure the polarized differential cross section,
\begin{eqnarray}
\frac{d \sigma_{pol}}{d {\rm cos}(\theta)} &=& \frac{1}{256 \pi s} \bigg[ \left(1+ \xi_{c}(x_1) \xi_{c}(x_2) \right) \left(  \mid {\cal M}_{++} \mid^2 + \mid {\cal M}_{--} \mid^2 \right) \nonumber \\ & & + \left(1- \xi_{c}(x_1) \xi_{c}(x_2) \right) \left(  \mid {\cal M}_{+-} \mid^2 + \mid {\cal M}_{-+} \mid^2 \right) \bigg] \, .
\end{eqnarray}

The anomalous structures relevant for the interaction $\gamma \gamma \rightarrow ZZ$ can be found in eq. (20).  We will write them in the now familiar form
\begin{eqnarray}
 {\cal O}_{1}^{Z Z} &=& \frac{e^2}{(\Lambda_{1})^4} \, ( F_{\mu \nu}F^{\mu \nu} ) ( Z^{\rho \sigma} \stackrel{\sim}{Z}_{\rho \sigma} ) \nonumber \\
 {\cal O}_{2}^{Z Z} &=& \frac{e^2}{(\Lambda_{2})^4} \, ( Z_{\mu \nu}Z^{\mu \nu} ) ( F^{\rho \sigma} \stackrel{\sim}{F}_{\rho \sigma} ) \nonumber \\
{\cal O}_{3}^{Z Z} &=& \frac{e^2}{(\Lambda_{3})^4} \, ( F_{\mu \nu}Z^{\mu \nu} ) ( F^{\rho \sigma} \stackrel{\sim}{Z}_{\rho \sigma} ) \nonumber \\
{\cal O}_{4}^{Z Z} &=& \frac{e^2}{(\Lambda_{4})^4} \, F^{\mu \nu} Z_{\nu \rho} F^{\rho \sigma} \stackrel{\sim}{Z}_{\sigma \mu} \nonumber \\
{\cal O}_{5}^{Z Z} &=& \frac{e^2}{(\Lambda_{5})^4} \, Z^{\mu \nu} F_{\nu \rho} Z^{\rho \sigma} \stackrel{\sim}{F}_{\sigma \mu} \,\, .
\end{eqnarray}
We have again set $g_{eff}=1$, where $g_{eff}$ is the coupling constant associated with each operator.  The detailed amplitudes are presented in the Appendix.  As in $\gamma \gamma \rightarrow \gamma Z$, those with one longitudinal $Z$ are suppressed by a factor of $m_{Z} / \sqrt{s}$, and we will not attempt to reconstruct them.  Amplitudes with two longitudinal $Z$s are $O(m_{Z}^2 /s)$, and hence negligible in our approximation, which again entails the use of the asymptotic SM expressions presented in~\cite{Goun_gg4}.  The amplitudes ${\cal M}^{CP}_{\pm \pm \pm \pm}$ vanish for ${\cal O}_{3}^{Z Z}$, and therefore this operator does not contribute to $A_{ZZ}$.   

The experimental reconstruction of this process is more difficult than for $\gamma \gamma \rightarrow \gamma \gamma$ or $\gamma \gamma \rightarrow \gamma Z$, as there is no high $p_T$ photon to tag.  Event reconstruction will likely require demanding a leptonic $Z$ decay or possibly a $Z \rightarrow b \bar{b}$ decay for one of the $Z$ bosons, resulting in a detection efficiency of roughly $40 \sim 45$\%.  However, it is beyond the scope of this paper to present a detailed $ZZ$ reconstruction analysis; such a study should also make use of the complete SM amplitudes and detector efficiencies.  An appropriate scaling of the integrated luminosity when observing the quoted sensitivities is sufficient to estimate the effects of imperfect reconstruction.

The binned asymmetries for the symmetric initial polarizations of ${\cal O}_{1}^{ZZ}$ are shown in fig. (7).  They are typically larger than those for $\gamma \gamma \rightarrow \gamma Z$.  Note that the central value of the $(+---)$ asymmetry displays a peculiar peaked shape.  The integrated asymmetries and total cross sections for the same initial polarization states are shown in fig. (8).  Again, the integrated asymmetry becomes large at low energies for all initial polarization states, suggesting that this asymmetry is a sensitive test of these anomalous operators.  The asymmetries for ${\cal O}_{2}^{ZZ}$ are of similar magnitude as those for ${\cal O}_{1}^{ZZ}$, but have the opposite sign.  Similarly, the asymmetries for ${\cal O}_{4}^{ZZ}$ and ${\cal O}_{5}^{ZZ}$ possess opposite signs, and are somewhat smaller in magnitude than those for ${\cal O}_{1}^{ZZ}$.  SU(2) $\times$ U(1) operators whose coefficients $c^{1}_{\alpha}$, $c^{2}_{\alpha}$ have the same sign will exhibit destructive interference between ${\cal O}_{1}^{ZZ}$ and ${\cal O}_{2}^{ZZ}$, while those with coefficients of opposite signs will see constructive interference; the same interference pattern exists for the coefficients $c^{4}_{\alpha}$ and $c^{5}_{\alpha}$.

To convert these results into statements regarding the sensitivity of this process to the various SU(2) $\times$ U(1) operators we need the coefficients $c^{i}_{\alpha}$ presented in table (2).
\begin{table}
\centering
\begin{tabular}{|l|l|l|l|l|l|} \hline\hline
$c^{i}_{\alpha}$ & $ {\cal O}_{1}^{ZZ}$ & $ {\cal O}_{2}^{Z Z}$ & $ {\cal O}_{3}^{ZZ}$ & $ {\cal O}_{4}^{ZZ}$ & $ {\cal O}_{5}^{ZZ}$ \\ \hline ${\cal O}_{(BB)(BB)}$ & $s_{W}^2 c_{W}^2$ & $s_{W}^2 c_{W}^2$ & $4s_{W}^2 c_{W}^2$ & $0$ & $0$ \\  ${\cal O}_{(WW)(WW)}$  & $s_{W}^2 c_{W}^2$ & $s_{W}^2 c_{W}^2$ & $4s_{W}^2 c_{W}^2$ & $0$ & $0$ \\  ${\cal O}_{(BB)(WW)}$ & $c_{W}^4$ & $s_{W}^4$ & $-4 s_{W}^2 c_{W}^2$ & $0$ & $0$ \\  ${\cal O}_{(WW)(BB)}$  & $s_{W}^4$ & $c_{W}^4$ & $-4 s_{W}^2 c_{W}^2$ & $0$ & $0$ \\  ${\cal O}_{(BW)(BW)}$ & $-c_{W}^2 s_{W}^2$ & $-c_{W}^2 s_{W}^2 $ & $(c_{W}^2 -s_{W}^2 )^2$ & $0$ & $0$ \\  ${\cal O}_{BWBW}$ & $-\frac{1}{2} c_{W}^2 s_{W}^2$ & $-\frac{1}{2} c_{W}^2 s_{W}^2$ & $0$ & $c_{W}^4$ & $s_{W}^4$ \\  ${\cal O}_{WBWB}$ & $-\frac{1}{2} c_{W}^2 s_{W}^2$ & $-\frac{1}{2} c_{W}^2 s_{W}^2$ & $0$ & $s_{W}^4$ & $c_{W}^4$ \\ \hline\hline
\end{tabular}
\caption{Coefficients $c^{i}_{\alpha}$ relating the ${\cal O}_{\alpha}^{SU(2) \times U(1)}$ to the ${\cal O}_{i}^{Z Z}$. }
\end{table}
Unlike the situation in $\gamma \gamma \rightarrow \gamma Z$, we see that all of the interference effects are destructive, given the relative signs of the $c^{i}_{\alpha}$ for each operator.  However, since $c_{W}^4 \sim 0.59$ and $s_{W}^4 \sim 0.053$, a complete cancellation does not occur except for the three operators ${\cal O}_{(BB)(BB)}$, ${\cal O}_{(WW)(WW)}$, and ${\cal O}_{(BW)(BW)}$.  We have performed a combined least-squares fit to the normalized binned cross section, binned asymmetry, and total cross section, with $ \sqrt{s}=1000$ GeV, for the polarization state $(+-+-)$ to estimate the value of $\Lambda_{\alpha}$ that can be probed in this process at a photon collider.  The search reaches obtainable from the other symmetric polarization states are similar.  The results are presented in fig. (9).  We only display results for the 
operators ${\cal O}_{(BB)(BB)}$, ${\cal O}_{(BW)(BW)}$, ${\cal O}_{(BB)(WW)}$, and ${\cal O}_{BWBW}$; the sensitivity to ${\cal O}_{(BB)(BB)}$ and ${\cal O}_{(WW)(WW)}$ is identical at high energies.  Similarly, ${\cal O}_{(BB)(WW)}$ and ${\cal O}_{(WW)(BB)}$ differ only in the sign of their asymmetries, as do ${\cal O}_{BWBW}$ and ${\cal O}_{WBWB}$, and hence yield identical search reaches.  We see that overall, this process is not quite as sensitive to the anomalous operators as $\gamma \gamma \rightarrow \gamma Z$ due to the destructive interference arising from the SU(2) $\times$ U(1) embedding of the $\gamma \gamma ZZ$ vertex structures.  However, those operators for which the asymmetry does not vanish can still be constrained quite stringently.

In conclusion, the sensitivity of $A_{ZZ}$ to interference with the large SM amplitude ${\cal M}_{\pm \pm \pm \pm}$ renders this process a sensitive probe of $CP$ violation appearing in gauge boson self couplings.

\noindent
\begin{figure}[htbp]
\centerline{
\psfig{figure=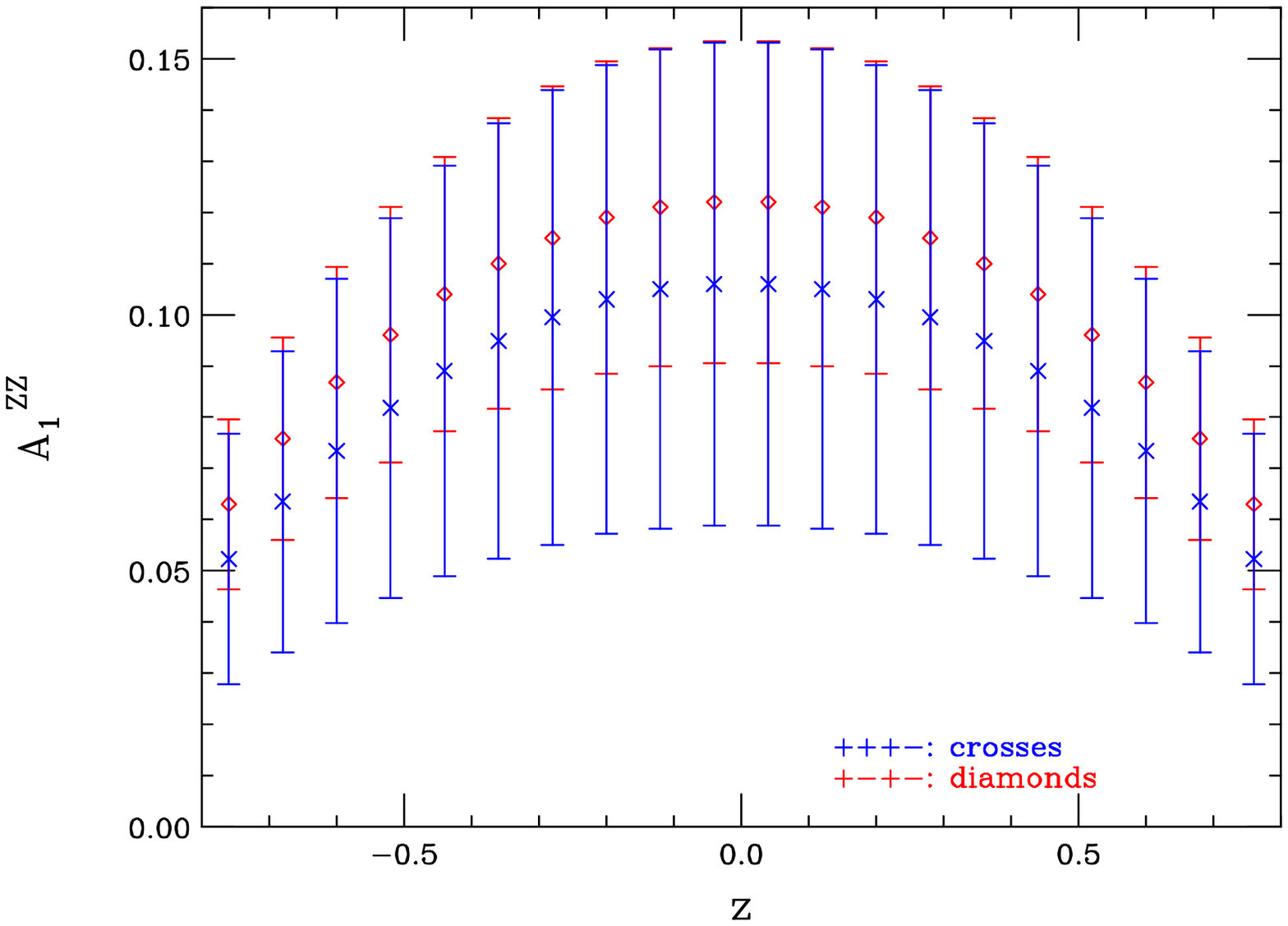,height=8.4cm,width=12.8cm,angle=0}}
\vspace*{1.0cm}
\centerline{
\psfig{figure=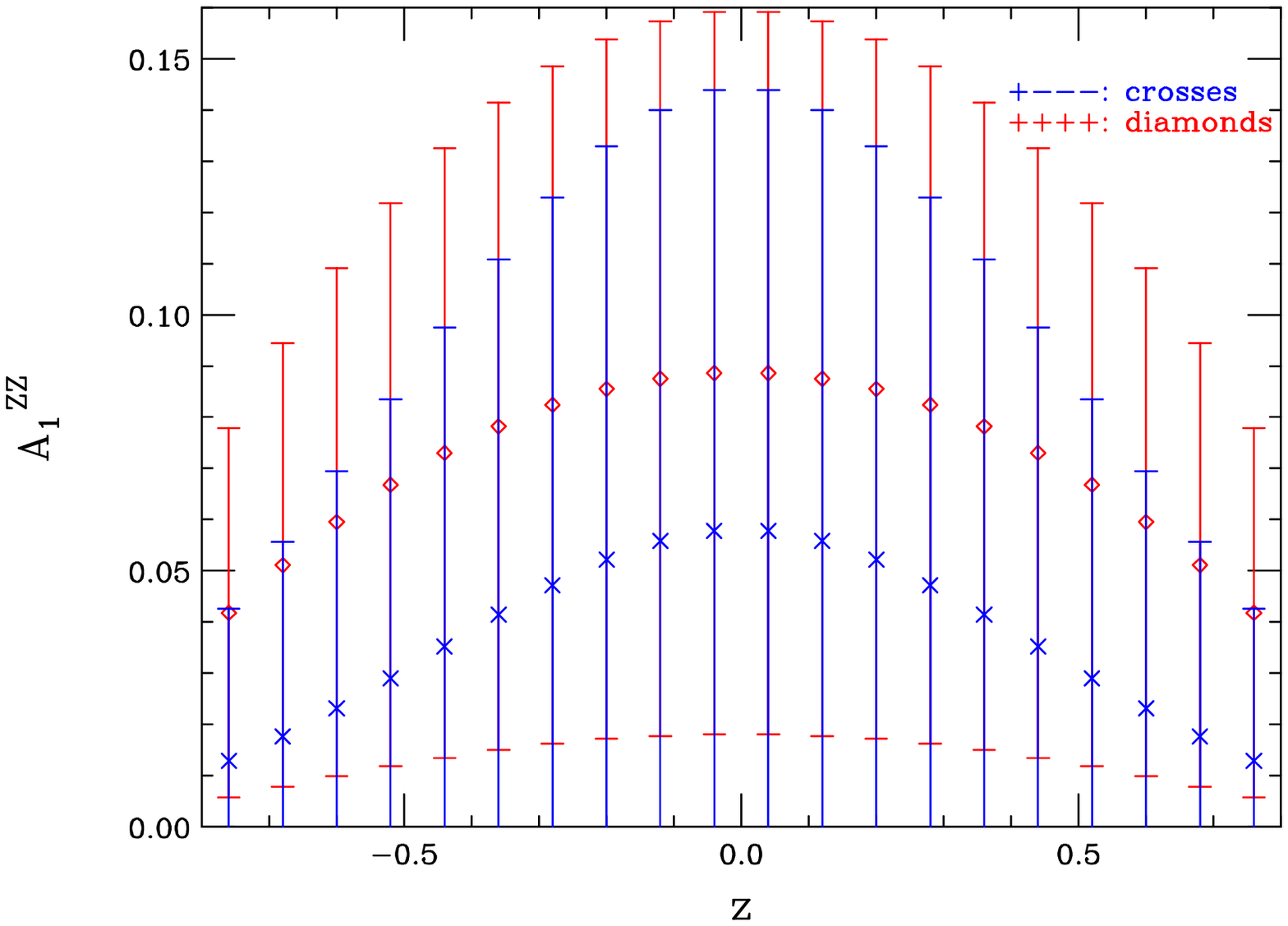,height=8.4cm,width=12.8cm,angle=0}}
\caption{Binned asymmetries for symmetric initial polarizations for ${\cal O}_{1}^{Z Z}$, with $\Lambda_1 = 2$ TeV, $L=500 \, {\rm fb^{-1}}$, and $\sqrt{s}=1000$ GeV.  The bars indicate the corresponding statistical error.}
\end{figure}

\noindent
\begin{figure}[htbp]
\centerline{
\psfig{figure=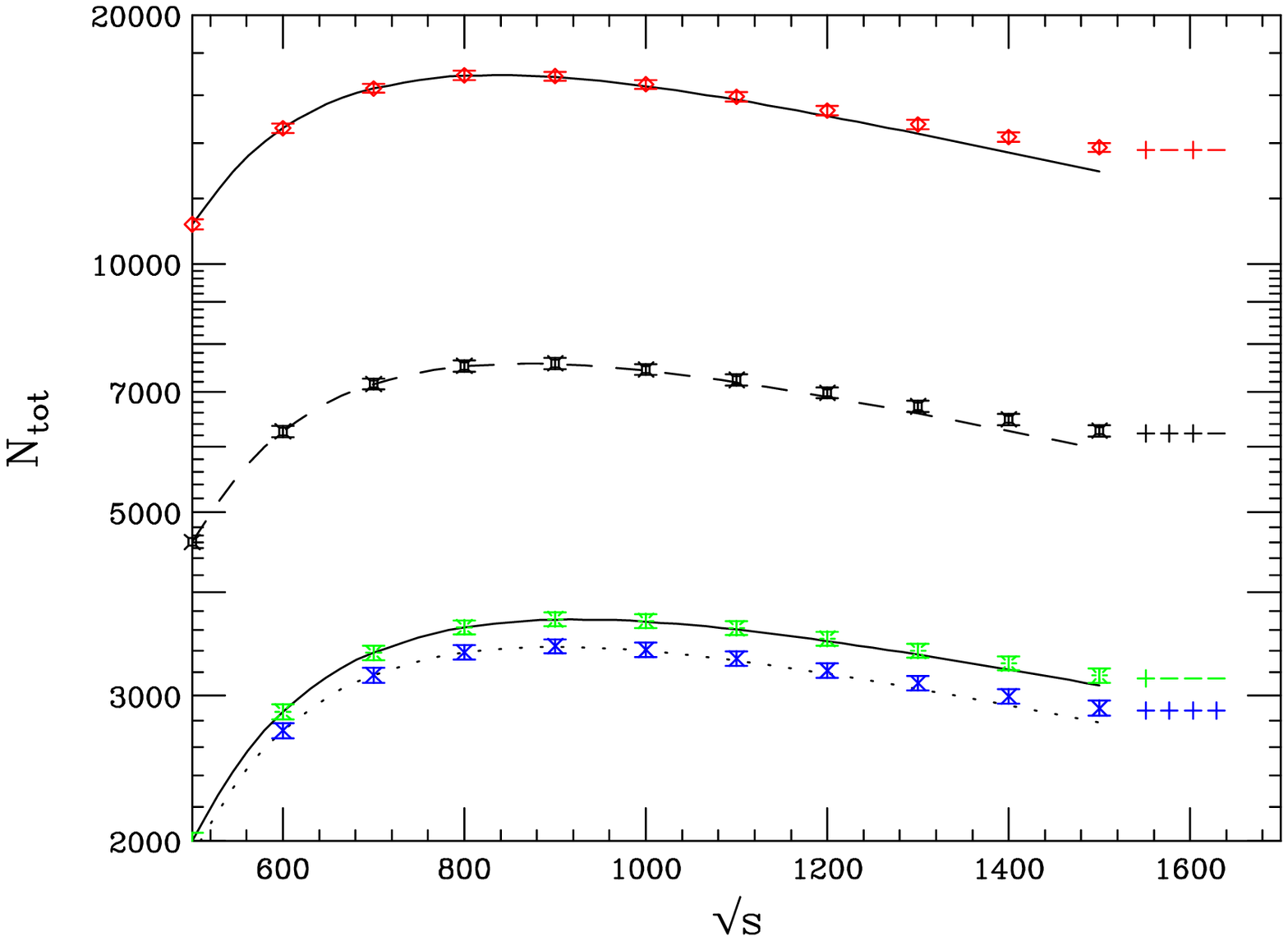,height=8.4cm,width=12.8cm,angle=0}}
\vspace*{1.0cm}
\centerline{
\psfig{figure=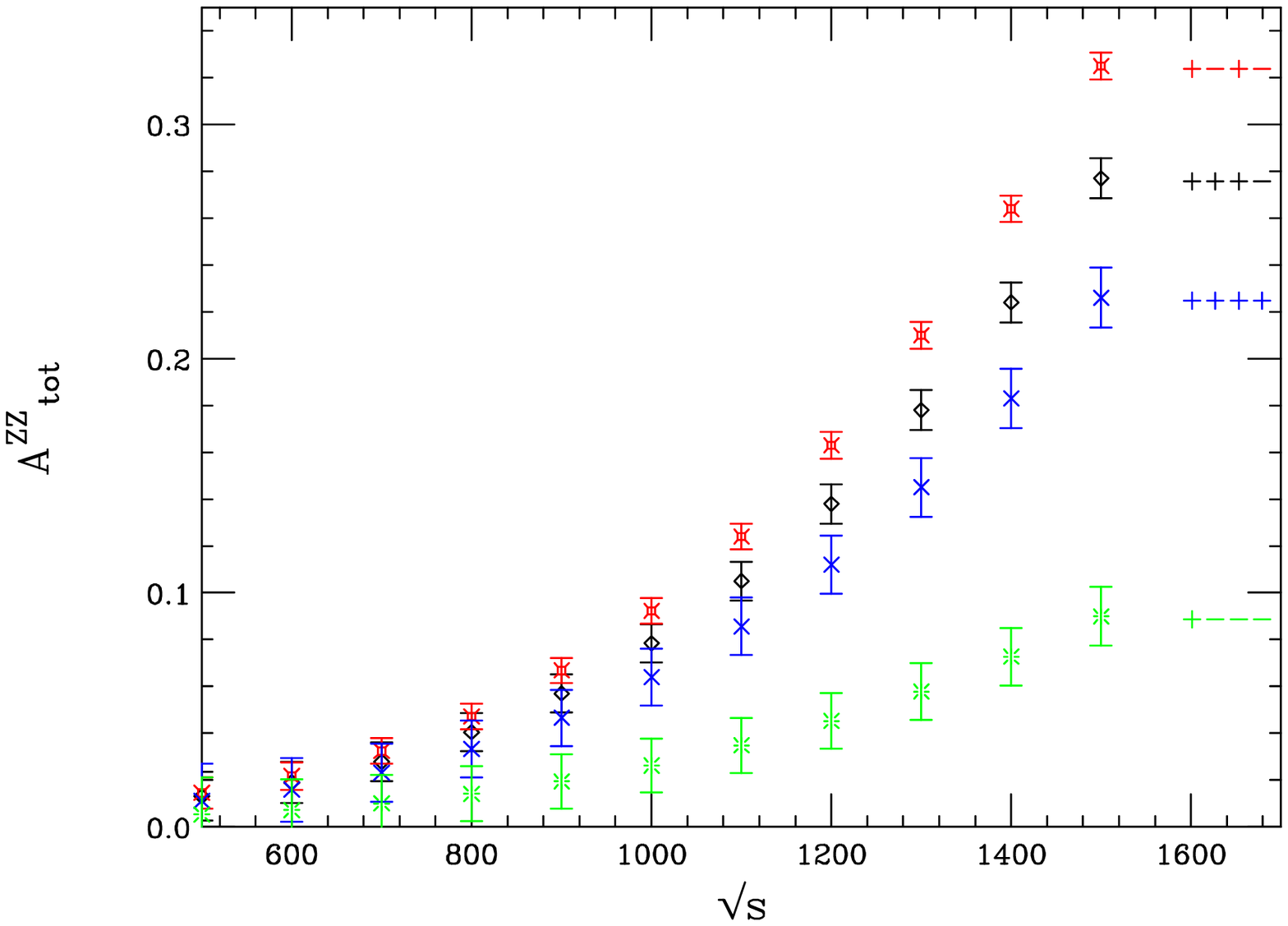,height=8.4cm,width=12.8cm,angle=0}}
\caption{Total (polarized) counting rate (top) and integrated asymmetry (bottom) versus $\sqrt{s}$ for ${\cal O}_{1}^{Z Z}$, with $\Lambda_1 = 2$ TeV and $L=500 \, {\rm fb^{-1}}$.  The bars indicate the corresponding statistical error, as well as a 1 \% luminosity uncertainty in the event rate.  The solid curves represent the SM event rates.}
\end{figure}

\noindent
\begin{figure}[htbp]
\centerline{
\psfig{figure=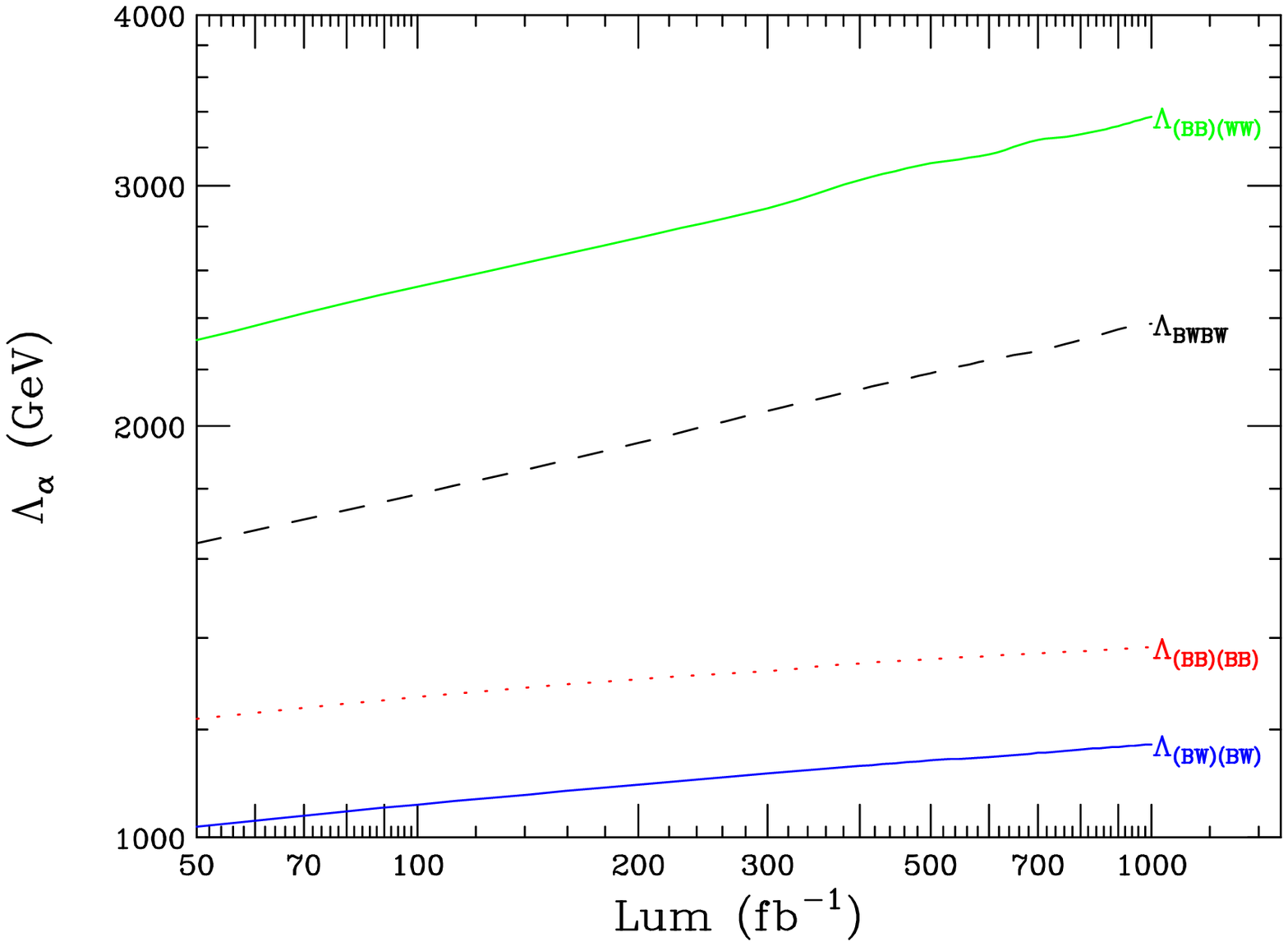,height=8.4cm,width=12.8cm,angle=0}}
\caption{Sensitivity to $\Lambda_{\alpha}$ for the polarization state $(+-+-)$, at the 95 \% CL.}
\end{figure}

\section{Conclusion}

In this paper we have examined the possibility of using the processes $\gamma \gamma \rightarrow \gamma \gamma$, $\gamma Z$, and $ZZ$ to search for sources of $CP$ violation that manifest themselves at energy scales above those of future colliders.  We have parameterized these effects by an effective Lagrangian containing the complete and independent set of $CP$-odd, SU(2) $\times$ U(1) invariant quartic gauge boson operators relevant to these interactions.  The considered processes vanish at tree level in the SM, and at one loop are completely dominated by the imaginary parts of the amplitudes ${\cal M}_{\pm \pm \pm \pm}$, ${\cal M}_{\pm \mp \pm \mp}$, and ${\cal M}_{\pm \mp \mp \pm}$.  The fact that they might be powerful tools in indirect searches for new physics was pointed out in~\cite{Goun_gg1,Goun_gg2,Goun_gg3,Goun_gg4}, where their sensitivity to supersymmetric loop effects was discussed.  Our results show that they are also of utility in searches for sources of $CP$ violation that contribute to self interactions of gauge bosons.

To demonstrate that these processes are indeed sensitive probes of the gauge boson sector, we compare our results with four others found in the literature.  We first note that while our operators are of dimension eight, those with which we compare are of dimension six.  Although our comparisons are for the search reaches for the new physics scale $\Lambda$, the reader should be aware that a given limit on $\Lambda$ translates into a greater sensitivity to a dimension eight operator, which scales as $s^2 / \Lambda^4$, than a dimension six operator, which scales as $s / \Lambda^2$, for $\sqrt{s} < \Lambda$.  We also note that the constraints on the anomalous SU(2) $\times$ U(1) operators obtained in this paper are stronger than those implied by unitarity.  We first
contrast our search reaches with the direct bounds on quartic gauge boson couplings given recently by the LEP experiments~\cite{LEP}.  They consider operators of the form $(e^2 a_{i}/16 \Lambda^2) \, {\cal O}$, and obtain bounds of $a_{i}/16 \Lambda^2 < 10^{-3} \, {\rm GeV}^2$ on $CP$-conserving $W^+ W^- \gamma \gamma$, $W^+ W^- Z \gamma$, and $ZZ\gamma \gamma$ vertices.  The operators considered in this paper are of the form $(e^2 / \Lambda^4) \, {\cal O}$ (with $g_{eff}=1$), and therefore our search reach for $1 / \Lambda^4 $ are to be contrasted with their bounds on $a_{i}/16 \Lambda^2$.  As expected, our sensitivity at a higher energy and higher luminosity collider are very significantly better.  We next compare our results with those of a study of these same operators at a $500$ GeV $e^+ e^-$ collider with an integrated luminosity of $500 \, {\rm fb}^{-1}$~\cite{Belanger}.  In this study the operators are parameterized in the form $(\kappa e^4)/(4 s^{2}_{W} c^{2}_{W} \Lambda^2) \, {\cal O}$, and bounds of $(\kappa e^2)/(4 s^{2}_{W} c^{2}_{W} \Lambda^2) < 0.13 \times 10^{-5} \, {\rm GeV}^2 $ obtained on vertices such as the $ZZ \gamma \gamma $ vertex.  Setting $\kappa =1$, this 
results in $\Lambda > 800 \sim 900$ GeV, at the $3 \sigma$ level.  The sensitivities obtained in this paper, normalized to the appropriate luminosity and energy are $\Lambda > 800 \sim 1700$ GeV, where the better constraints are for those operators contributing to one of the considered asymmetries.  The search reaches obtained here on $\Lambda$, after adjusting the statistical significances, are still stronger, even though the operators we consider are of dimension eight.  The sensitivity of future hadron colliders such as the LHC to these anomalous $CP$-even $\gamma \gamma ZZ$ and $\gamma \gamma W^+ W^-$ has been studied in~\cite{Hadron}.  This paper obtains search reaches of $\Lambda > 800 \sim 1100$ at the 95 \% CL, which are again somewhat weaker than those we have presented.  Finally, the authors of~\cite{Choi} consider $CP$-odd dimension six operators contributing to the process $\gamma \gamma \rightarrow W^+ W^-$ at $\sqrt{s} = 1$ TeV assuming $20 \, {\rm fb}^{-1}$ of integrated luminosity.  They cast their operators in the form $(Y / m^{2}_{W}) \, {\cal O}$, and obtain bounds of $Y < 2 \times 10^{-3} \sim 4 \times 10^{-5}$.  Note the we have improved their results by a factor of 2.5 to simulate the search reach assuming $50 \, {\rm fb}^{-1}$ of integrated luminosity.  These translate into limits of $\Lambda > 500 \sim 4100$ GeV in our notation, at the $1 \sigma$ level.  The relevant limits in this paper to compare to are those 
assuming $50 \, {\rm fb}^{-1}$ of integrated luminosity, which range from $\Lambda > 800$ GeV to $\Lambda > 2500$ GeV at the 95 \% CL.  Our constraints on the new physics scale $\Lambda$ are not quite as strong, which is not too surprising considering the large cross section for $\gamma \gamma \rightarrow W^+ W^-$ and the fact that their operators are of dimension six while ours are of dimension eight, although the differences are lessened after accounting for the greater statistical significance of our results.  However, the processes considered are complementary; they study the charged gauge boson production process $\gamma \gamma \rightarrow W^+ W^-$ while we examine the neutral gauge boson processes $\gamma \gamma \rightarrow \gamma \gamma$, $\gamma Z$, and $ZZ$.   

In summary, the processes  $\gamma \gamma \rightarrow \gamma \gamma$, $\gamma Z$, and $ZZ$ are sensitive probes of $CP$ violation in the gauge boson sector.  The sensitivities obtained here to the anomalous $CP$-odd operators affecting these interactions compare favorably with similar limits on anomalous gauge boson couplings found in the literature.  In addition, the examination of these processes nicely complements previous studies that have focused primarily on $W$ boson, top quark, or Higgs production.  Their utility in other indirect physics searches at future photon colliders should certainly be studied in more detail.

\noindent{\Large\bf Acknowledgements}

It is a pleasure to thank Hooman Davoudiasl for several useful conversations, and Tom Rizzo for suggesting this project and for many helpful discussions.  The work of F. P. was supported in part by an NSF Graduate Research Fellowship.

\noindent{\Large\bf Appendix}

We collect here formulae describing the various photon distribution functions; derivations are given in~\cite{Russians}.  We first define the auxiliary functions
\begin{equation}
C(x) \equiv \frac{1}{1 - x} + (1 - x) - 4 r (1 - r) - P_e \, P_l \, r \, z 
(2 r - 1) (2 - x),
\label{C(x)}
\end{equation}
where $r = x/[z(1 - x)]$, and
\[
\sigma_{_{C}} = \left(\frac{2 \pi \alpha^2}{m_e^2 z}\right) \left[\left(1 - 
\frac{4}{z} -\frac{8}{z^2}\right) \ln (z + 1) + \frac{1}{2} + \frac{8}{z} - 
\frac{1}{2 (z + 1)^2}\right]\]
\begin{equation}
+ P_e \, P_l \left(\frac{2 \pi \alpha^2}{m_e^2 z}\right) \left[\left(1 + 
\frac{2}{z}\right) \ln (z + 1) - \frac{5}{2} + \frac{1}{z + 1} - 
\frac{1}{2 (z + 1)^2}\right].  
\end{equation}
Here $z$ is a variable describing the laser photon energy, and is given by $z=4 E_{e}E_{l}/m^{2}_{e}$, where $E_{e}$ is the electron beam energy and $E_{l}$ the laser energy.  Varying $z$ changes the maximum value of the backscattered photon beam energy $x_{max}$, where $x_{max}=z/(1+z)$.  We will set $z=2(1+\sqrt{2})$, which maximizes $x_{max}$ while preventing interactions between the backscattered photons and laser beam.  In terms of these functions and variables the photon number and 
helicity distributions take the form
\begin{eqnarray}
f(x, P_e, P_l; z) &=& \left(\frac{2 \pi \alpha^2}{m_e^2 z 
\sigma_{_{C}}}\right) C(x)  \nonumber \\ 
\xi_{c}(x, P_e, P_l; z) &=& \frac{1}{C(x)}\left\{P_e \, \left[\frac{x}{1 - x} 
+ x (2 r - 1)^2\right]
 - P_l \, (2r - 1)\left(1 - x + \frac{1}{1 - x}\right)\right\}\,. \nonumber \\
\xi_{t}(x,P_e, P_l, P_t; z) &=& \frac{2 r^2 P_t }{C(x)} \,\,.
\end{eqnarray}
$f$ is the photon number density function, while $\xi_c $ and $\xi_t $ are respectively the circular and linear helicity distribution functions.  We can now write the observable differential cross section as
\begin{equation}
\frac{d \sigma}{d \Omega} =  \frac{1}{64 \pi^2 n!} \, \int \int \, dx_1 dx_2 \frac{f(x_1)f(x_2)}{x_1 x_2 s} \, \mid {\cal M} \mid^{2}_{ens} \,\, ,
\end{equation}
where $s=4E^{2}_{e}$ and $n!$ accounts for any final state Bose symmetries.

Here we present the kinematics and amplitudes used for the processes in this paper.  The generic interaction considered is $F_{1}(k_1, \epsilon_1) + F_2(k_2, \epsilon_2) \rightarrow V_1(p_1, V_1) + V_2(p_2, V_2)$, where $F_1$, $F_2$ are the incoming photons and $V_1$, $V_2$ the outgoing gauge bosons.  The momenta and polarization vectors take the following form in the c.m.s. frame:
\begin{equation}
\begin{array}{rcl}
k_{1\mu} &=& {\sqrt{\hat{s}}\over {2}}(1,0,0,1)\\
p_{1\mu} &=& {\sqrt{\hat{s}}\over {2}}(1,\beta s_\theta,0,\beta c_\theta)\\ 
\epsilon_1^\mu &=& -{1\over {\sqrt 2}}(0,\lambda_1,i,0)\\
V_{1L}^{\mu *} &=& {\sqrt{\hat{s}}\over {2m}}(-\beta,s_\theta,0,c_\theta)\\
V_{1T}^{\mu *} &=& {1\over {\sqrt 2}}(0,-\lambda_3 c_\theta,i,\lambda_3
s_\theta)
\end{array}\qquad
\begin{array}{rcl}
k_{2\mu} &=& {\sqrt{\hat{s}}\over {2}}(1,0,0,-1)\\
p_{2\mu} &=& {\sqrt{\hat{s}}\over {2}}(1,-\beta s_\theta,0,-\beta c_\theta)\\ 
\epsilon_2^\mu &=& {1\over {\sqrt 2}}(0,-\lambda_2,i,0)\\
V_{2L}^{\mu *} &=& -{\sqrt{\hat{s}}\over {2m}}(\beta,s_\theta,0,c_\theta)\\
V_{2T}^{\mu *} &=& {1\over {\sqrt 2}}(0,-\lambda_4 c_\theta,-i,\lambda_4
s_\theta)
\end{array}
\end{equation}
with $s_\theta=\sin \theta$, $c_\theta=\cos \theta$, $\beta_V=\sqrt{1-4m_{V}^2 / \hat{s}}$, and $\hat{s}=(k_1 + k_2)^2 = (p_1 + p_2)^2$.  Given these expressions, it is straightforward to compute the amplitudes for the operators of eq. (20).  As mentioned in the text we neglect terms of $O(m_{W}^2 / \hat{s}, m_{Z}^2 / \hat{s})$, as the SM amplitudes we use also omit terms of this order.  We find that this approximation is numerically valid for the energies considered here.  The amplitude for the anomalous $\gamma \gamma$ operator of eq. (20) takes the form
\begin{equation}
{\cal M}_{1}^{\gamma \gamma}(\lambda_1,\lambda_2;\lambda_3,\lambda_4) = \frac{ie^2 \hat{s}^2}{2(\Lambda_{1}^{\gamma \gamma})^4} \, \bigg[ (\lambda_3 + \lambda_4)(1+\lambda_1 \lambda_2 ) - (\lambda_1 + \lambda_2)(1+\lambda_3 \lambda_4) \bigg] \left(3+c^{2}_{\theta} \right ) \, .
\end{equation}
This amplitude is exact, as $m_Z$ and $m_W$ appear only in the SM amplitudes.  Note that this amplitude is odd under the interchange $(\lambda_1 , \lambda_2 ) \leftrightarrow (\lambda_3 , \lambda_4 )$; this is a consequence of $T$ violation and the fact that the two final state particles are the same as the initial states.  The anomalous amplitudes for the process $\gamma \gamma \rightarrow \gamma Z$ (with $V_2 = Z$) are, neglecting terms of $O(m_{Z}^2 / \hat{s})$, 
\begin{eqnarray}
{\cal M}_{1}^{\gamma Z}(\lambda_1,\lambda_2;\lambda_3,\lambda_4) &=& \frac{ie^2 \hat{s}^2}{8(\Lambda_{1}^{\gamma Z})^4} \, \bigg[ 4(\lambda_3 + \lambda_4)(1+\lambda_1 \lambda_2 ) + 2(\lambda_4 + \lambda_1 \lambda_2 \lambda_3 ) - (\lambda_1 + \lambda_2 )(1+\lambda_3 \lambda_4 ) \nonumber \\ & & +2(\lambda_2 - \lambda_1)(1-\lambda_3 \lambda_4 )c_{\theta} + \bigg(2 (\lambda_4 + \lambda_1 \lambda_2 \lambda_3 ) - (\lambda_1 + \lambda_2 )(1+ \lambda_3 \lambda_4 ) \bigg) c^{2}_{\theta} \bigg] \nonumber \\
{\cal M}_{1}^{\gamma Z}(\lambda_1,\lambda_2;\lambda_3,0) &=& \frac{ie^2 m_Z \hat{s}}{2(\Lambda_{1}^{\gamma Z})^4} \, s_{\theta} \, \sqrt{\frac{\hat{s}}{2}} \, \bigg[\lambda_3 (\lambda_1 - \lambda_2) +\bigg(2- \lambda_3 (\lambda_1 +\lambda_2)\bigg) c_{\theta} \bigg]  \nonumber \\
{\cal M}_{2}^{\gamma Z}(\lambda_1,\lambda_2;\lambda_3,\lambda_4) &=& \frac{ie^2 \hat{s}^2}{8(\Lambda_{2}^{\gamma Z})^4} \, \bigg[ -5(\lambda_1 + \lambda_2)(1+\lambda_3 \lambda_4 ) + 2(\lambda_3 + \lambda_1 \lambda_2 \lambda_4 ) \nonumber \\ & & -2(\lambda_2 - \lambda_1)(1-\lambda_3 \lambda_4 )c_{\theta} + \bigg(2 (\lambda_3 + \lambda_1 \lambda_2 \lambda_4 ) - (\lambda_1 + \lambda_2 )(1+ \lambda_3 \lambda_4 ) \bigg) c^{2}_{\theta} \bigg] \nonumber \\
{\cal M}_{2}^{\gamma Z}(\lambda_1,\lambda_2;\lambda_3,0) &=& \frac{ie^2 m_Z \hat{s}}{2(\Lambda_{2}^{\gamma Z})^4} \, s_{\theta} \, \sqrt{\frac{\hat{s}}{2}} \, \bigg[ \lambda_3 (\lambda_2 - \lambda_1 ) + \bigg(2 \lambda_1 \lambda_2 - \lambda_3 (\lambda_1 + \lambda_2 ) \bigg) c_{\theta} \bigg] \nonumber \\
{\cal M}_{3}^{\gamma Z}(\lambda_1,\lambda_2;\lambda_3,\lambda_4) &=& \frac{ie^2 \hat{s}^2}{32(\Lambda_{3}^{\gamma Z})^4} \, \bigg[ 3(\lambda_1 + \lambda_2)(1+\lambda_3 \lambda_4 ) + 12(\lambda_4 + \lambda_1 \lambda_2 \lambda_3 ) +6(\lambda_3 + \lambda_1 \lambda_2 \lambda_4 ) \nonumber \\ & & +6(\lambda_2 - \lambda_1)(1-\lambda_3 \lambda_4 )c_{\theta} + \bigg(2 (\lambda_4 + \lambda_1 \lambda_2 \lambda_3 ) - (\lambda_1 + \lambda_2 )(1+ \lambda_3 \lambda_4 )  \nonumber \\ & & + (\lambda_4 - \lambda_3 )(1- \lambda_1 \lambda_2 ) \bigg) c^{2}_{\theta} \bigg] \nonumber \\
{\cal M}_{3}^{\gamma Z}(\lambda_1,\lambda_2;\lambda_3,0) &=& \frac{ie^2 m_Z \hat{s}}{8(\Lambda_{3}^{\gamma Z})^4} \, s_{\theta} \, \sqrt{\frac{\hat{s}}{2}} \, \bigg[ \lambda_3 (\lambda_1 - \lambda_2 ) + \bigg( 4 -2 \lambda_1 \lambda_2 - \lambda_3 (\lambda_1 + \lambda_2 ) \bigg) c_{\theta} \bigg] \, .
\end{eqnarray}
Again neglecting terms of $O(m_{Z}^2 / \hat{s})$, the $ZZ$ amplitudes become
\begin{eqnarray}
{\cal M}_{1}^{Z Z}(\lambda_1,\lambda_2;\lambda_3,\lambda_4) &=& \frac{ie^2 \hat{s}^2}{(\Lambda_{1}^{Z Z})^4} \, \left(\lambda_3 + \lambda_4 \right) \left(1+ \lambda_1 \lambda_2 \right) \nonumber \\
{\cal M}_{2}^{Z Z}(\lambda_1,\lambda_2;\lambda_3,\lambda_4) &=& -\frac{ie^2 \hat{s}^2}{(\Lambda_{2}^{Z Z})^4} \, \left(\lambda_1 + \lambda_2 \right) \left(1+ \lambda_3 \lambda_4 \right) \nonumber \\
{\cal M}_{3}^{Z Z}(\lambda_1,\lambda_2;\lambda_3,\lambda_4) &=& \frac{ie^2 \hat{s}^2}{8(\Lambda_{3}^{Z Z})^4} \, \bigg[ (\lambda_3 + \lambda_4)(1+\lambda_1 \lambda_2 ) - (\lambda_1 + \lambda_2)(1+\lambda_3 \lambda_4) \bigg] \left(1+c^{2}_{\theta} \right ) \nonumber \\
{\cal M}_{3}^{Z Z}(\lambda_1,\lambda_2;\lambda_3,0) &=& \frac{ie^2 m_Z \hat{s}}{2(\Lambda_{3}^{Z Z})^4} \, s_{\theta} \, \sqrt{\frac{\hat{s}}{2}} \, \bigg[ (1+ \lambda_1 \lambda_2 ) - \lambda_3 (\lambda_1 + \lambda_2 ) \bigg] c_{\theta} \nonumber \\
{\cal M}_{4}^{Z Z}(\lambda_1,\lambda_2;\lambda_3,\lambda_4) &=& \frac{ie^2 \hat{s}^2}{32(\Lambda_{4}^{Z Z})^4} \, \bigg[ 6 (\lambda_1 + \lambda_2 )(1+ \lambda_3 \lambda_4 ) + 2 (\lambda_3 + \lambda_4 )(1+ \lambda_1 \lambda_2 ) \nonumber \\ & & + 2 \bigg( - (\lambda_1 + \lambda_2 )(1+ \lambda_3 \lambda_4 ) + (\lambda_3 + \lambda_4 )(1+ \lambda_1 \lambda_2 ) \bigg) c^{2}_{\theta} \bigg] \nonumber \\
{\cal M}_{4}^{Z Z}(\lambda_1,\lambda_2;\lambda_3,0) &=& \frac{ie^2 m_Z \hat{s}}{4(\Lambda_{4}^{Z Z})^4} \, s_{\theta} \, \sqrt{\frac{\hat{s}}{2}} \, \bigg[ (1+ \lambda_1 \lambda_2 ) - \lambda_3 (\lambda_1 + \lambda_2 ) c_{\theta} \bigg] \nonumber \\
{\cal M}_{5}^{Z Z}(\lambda_1,\lambda_2;\lambda_3,\lambda_4) &=& \frac{ie^2 \hat{s}^2}{32(\Lambda_{5}^{Z Z})^4} \, \bigg[ -6 (\lambda_3 + \lambda_4 )(1+ \lambda_1 \lambda_2 ) - 2 (\lambda_1 + \lambda_2 )(1+ \lambda_3 \lambda_4 ) \nonumber \\ & & + 2 \bigg( (\lambda_3 + \lambda_4 )(1+ \lambda_1 \lambda_2 ) - (\lambda_1 + \lambda_2 )(1+ \lambda_3 \lambda_4 ) \bigg) c^{2}_{\theta} \bigg] \nonumber \\
{\cal M}_{5}^{Z Z}(\lambda_1,\lambda_2;\lambda_3,0) &=& \frac{ie^2 m_Z \hat{s}}{4(\Lambda_{5}^{Z Z})^4} \, s_{\theta} \, \sqrt{\frac{\hat{s}}{2}} \, \bigg[ (1+ \lambda_1 \lambda_2 ) - \lambda_3 (\lambda_1 + \lambda_2 ) c_{\theta} \bigg]
\end{eqnarray}
Amplitudes with two longitudinal $Z$ bosons are of $O(m_{Z}^2 / \hat{s})$, and hence are negligible in our approximation.

\end{document}